\documentclass[a4paper,10pt]{article}
\usepackage[utf8]{inputenc}
 \usepackage{slashed} 
   \usepackage{amsmath}
\usepackage[a4paper,bindingoffset=0.2in,left=1in,right=1in,top=1in,bottom=1in,footskip=.25in]{geometry}
\usepackage{graphicx,subfigure,epsfig,epsf}
\newcommand{\beq}{\begin{equation}}
\newcommand{\eeq}{\end{equation}}
\newcommand{\bea}{\begin{eqnarray}}
\newcommand{\eea}{\end{eqnarray}}
\newcommand{\bi}{\begin{itemize}}
\newcommand{\ei}{\end{itemize}}

\newcommand{\Lda}{\Lambda}
\newcommand{\g}{\gamma}
\def\Re{{\cal R \mskip-4mu \lower.1ex \hbox{\it e}\,}}
\def\Im{{\cal I \mskip-5mu \lower.1ex \hbox{\it m}\,}}

\def\etal{{\it et al.~}}

\def\tev{\,{\ifmmode\mathrm {TeV}\else TeV\fi}}
\def\gev{\,{\ifmmode\mathrm {GeV}\else GeV\fi}}
\def\mev{\,{\ifmmode\mathrm {MeV}\else MeV\fi}}
\def\to{\rightarrow}

\usepackage{color}

\begin{document}

% Reference Macros:  Enter parameters in order Vol, Page, Year
\def\issue(#1,#2,#3){{\bf #1}, #2 (#3)} % AIP format

\def\APP(#1,#2,#3){Acta Phys.\ Polon.\ \issue(#1,#2,#3)}
\def\ARNPS(#1,#2,#3){Ann.\ Rev.\ Nucl.\ Part.\ Sci.\ \issue(#1,#2,#3)}
\def\CPC(#1,#2,#3){Comp.\ Phys.\ Comm.\ \issue(#1,#2,#3)}
\def\CIP(#1,#2,#3){Comput.\ Phys.\ \issue(#1,#2,#3)}
\def\EPJC(#1,#2,#3){Eur.\ Phys.\ J.\ C\ \issue(#1,#2,#3)}
\def\EPJD(#1,#2,#3){Eur.\ Phys.\ J. Direct\ C\ \issue(#1,#2,#3)}
\def\IEEETNS(#1,#2,#3){IEEE Trans.\ Nucl.\ Sci.\ \issue(#1,#2,#3)}
\def\IJMP(#1,#2,#3){Int.\ J.\ Mod.\ Phys. \issue(#1,#2,#3)}
\def\JHEP(#1,#2,#3){J.\ High Energy Physics \issue(#1,#2,#3)}
\def\JMP(#1,#2,#3){J.\ Math.\ Phys.\ \issue(#1,#2,#3)}
\def\JPG(#1,#2,#3){J.\ Phys.\ G \issue(#1,#2,#3)}
\def\MPL(#1,#2,#3){Mod.\ Phys.\ Lett.\ \issue(#1,#2,#3)}
\def\NP(#1,#2,#3){Nucl.\ Phys.\ \issue(#1,#2,#3)}
\def\NIM(#1,#2,#3){Nucl.\ Instrum.\ Meth.\ \issue(#1,#2,#3)}
\def\PL(#1,#2,#3){Phys.\ Lett.\ \issue(#1,#2,#3)}
\def\PRD(#1,#2,#3){Phys.\ Rev.\ D \issue(#1,#2,#3)}
\def\PRL(#1,#2,#3){Phys.\ Rev.\ Lett.\ \issue(#1,#2,#3)}
\def\PTP(#1,#2,#3){Progs.\ Theo.\ Phys. \ \issue(#1,#2,#3)}
\def\RMP(#1,#2,#3){Rev.\ Mod.\ Phys.\ \issue(#1,#2,#3)}
\def\SJNP(#1,#2,#3){Sov.\ J. Nucl.\ Phys.\ \issue(#1,#2,#3)}

%%%%%%%%%%%%%%%%%%%%%%%%%%%%%
\title{Search for associated production of Higgs with $Z$ boson in the noncommutative Standard Model at linear colliders}
\author{Selvaganapathy J$^{1}$, Prasanta Kumar Das$^{1}$ and Partha Konar$^{2}$ \\
{\it $^{1}$ Department of Physics, Birla Institute of Technology and Science-Pilani,} \\
{\it K K Birla Goa campus, Goa-403726, India} \\
{\it $^{2}$Physical Research Laboratory, Ahmedabad-380009, India} \\
}

\maketitle

\begin{abstract}
We study the associated Higgs production with $Z$ boson  at future linear colliders in the framework of the minimal noncommutative standard model. Using the  Seiberg-Witten map, we calculate the production cross-section considering all orders of the noncommutative parameter $\Theta_{\mu\nu}$. We consider the effect of earth's rotation on the orientation of $\Theta_{\mu\nu}$ with respect to the laboratory frame and thus on the total cross-section, it's azimuthal distribution and rapidity distribution for the machine energy ranging from $0.5~\rm{TeV}$ to $3~\rm{TeV}$ corresponding to the noncommutative scale $\Lambda \ge 0.5~\rm{TeV}$. 
\end{abstract}

\newpage
\tableofcontents

\section{Introduction}
Recently, the models of extra-spatial dimension where the gravity is strong at the TeV scale \cite{ADD,AADD}, has drawn a lot of attention among the physics community. Thanks to the TeV scale new physics, even if some tiny amount of space-time noncommutativity is realised in nature, one can possibly consider the effects of this at the currently running multi-TeV energy hadron collider like LHC or the upcoming TeV energy electron-positron linear colliders. Much cleaner and hadronically quiet environment together with the complete knowledge on incoming beam momenta at the linear collider permits the complete reconstruction of events. Hence, one can expect a precision measurements of the effects from the new physics. Although the idea is rather old, lately remarkable interests arose in the noncommutative (NC) field theory from the pioneering work of Snyder \cite{Snyder}. 
%The advances in D-brane dynamics in string theory at low-energy reveals that space-time can possibly be noncommutative \cite{CDS,DH,SW,Witten,HW}. 

 The space and time coordinates at the TeV energy become operator satisfying
\beq \label{XXTheta}
[\hat{X}_\mu,\hat{X}_\nu]= i\Theta_{\mu\nu} = i\frac{c_{\mu\nu}}{\Lda^2}.
%\label{NCSTh}
\eeq
where $\Theta_{\mu\nu}$ is an antisymmetric matrix tensor and of dimension $[M]^{-2}$.
%It  has the dimension of area and it reflects the extent to which the spacetime is fuzzy i.e. noncommutative. 
$\Lda~$ is the NC scale at which the effect of space-time noncommutativity shows up. $c_{\mu\nu}$ is the anti-symmetric c-number matrix. 

There are several approaches to study the effect of space-time noncommutativity in a field theory. One is the Moyal-Weyl (MW) approach in which one replaces the ordinary product between two functions $\phi(x)$ and $\psi(x)$ in terms of $\star$ (Moyal-Weyl) product defined by a formal power series expansion of
%%%%%%%%%%%%%%%%%%%%%%%%%%%%%
\cite{RJ,JS,BV}
\begin{equation}
(f \star g)(x)=exp\left(\frac{1}{2}\Theta_{\mu\nu}\partial_{x^\mu}\partial_{y^\nu}\right)f(x)g(y)|_{y=x}.
\label{StarP}
\end{equation}
%%%%%%%%%%%%%%%%%%%%%%%%%%%%%
together with the  ordinary integral $\int d^n x f(x)$ which has the property
\bea
\int d^n x (f \star g)(x) = \int d^n x (g \star f)(x) = \int d^n x f(x) g(x)  
\eea
Here $f(x)$ and $g(x)$ are ordinary functions on $R^n$ and the expansion in the star product can be seen intuitively as an expansion of the product in its non-commutativity. 
 Extensive works were done on renormalisation of the NC field theory \cite{Filk,VB,CDP1,CDP2,VS,ABK}. The non-trivial phase factor(arises after summing over all orders of $\Theta$ in the star product defined in the Weyl-Moyal plane) which gives rise the UV/IR mixing problem \cite{MVS,Raamsdonk}. Detailed collider searches of space-time noncommutativity include the following: Hewett \etal \cite{HPR,Hewett1,Hewett2} and others \cite{Mathews,Rizzo} investigated e.g. $e^+ e^- \to e^+ e^-$ (Bhabha), $e^- e^- \to e^- e^-$ (M\"{o}ller), $e^- \g \to e^- \g$, $e^+ e^- \to \g \g$ (pair annihilation), $\g \g \to e^+ e^-$ and $\g \g \to \g \g$ in the context of NCQED. For a review on the NC phenomenology see \cite{HKM}.

 Another approach is the Seiberg-Witten approach in which the space-time noncommutativity is being treated perturbatively via the Seiberg-Witten (SW) map expansion of the fields in terms of noncommutative parameter $\Theta$ \cite{DH,SW,Witten,HW}. Here the gauge parameter $\lambda$ and the gauge field $A^\mu$ is expanded as 
\bea \label{swps}
\lambda_\alpha (x,\Theta) &=& \alpha(x) + \Theta^{\mu\nu} \lambda^{(1)}_{\mu\nu}(x;\alpha) + \Theta^{\mu\nu} \Theta^{\eta\sigma} \lambda^{(2)}_{\mu\nu\eta\sigma}(x;\alpha) + \cdot \cdot \cdot \\
A_\rho (x,\Theta) &=& A_\rho(x) + \Theta^{\mu\nu} A^{(1)}_{\mu\nu\rho}(x) + \Theta^{\mu\nu} \Theta^{\eta\sigma} A^{(2)}_{\mu\nu\eta\sigma\rho}(x) + \cdot \cdot \cdot
\eea

The advantage in the SW aprroach over the Weyl-Moyal approach is that it can be applied to any gauge theory and matter can be in an arbitrary representation.  %Also note that the non-trivial phase factor(arises after summing over all orders of $\Theta$ in the star product defined in the Weyl-Moyal plane) which gives rise the UV/IR mixing, does not show up in the noncommutative gauge theories which uses the $\Theta$-expansion method of Seiberg-Witten, a theory which is often claimed to be free from UV/IR mixing \cite{MST}.
Reasonable progress has been made in NCQED using  SW map as far as it's quantum structure, perturbative renormalization is concerned
\cite{MST,BGGPSW,Martin,GV,BMR,BRT,SY,Horvat,HITY1,HISTY2,HITY3,TY}.
In particularly, the UV/IR mixing arised in an arbitrary non-abelian noncommutative gauge theory has been studied 
\cite{MST,BGGPSW,GV,SY,Horvat,HITY1,HISTY2,HITY3,TY}. Using SW expansion of the NC fields, Bichl \etal  ~ \cite{BGGPSW} show that the photon self-energy in NCQED is renormalizable to all orders in $\Theta$.  Anomalies and renormalizability of SW $\theta$-expanded NCSM are studied in \cite{Martin,BMR,BRT}. 
The SW map based $\theta$-exact model two-point functions for photon and neutrino are discussed in \cite{SY,HITY1,HISTY2,HITY3,TY}.
%and the way how to eliminate quantum pathologies was found.
%Martin \etal \cite{Martin} show that in any GUT inspired noncommutative theory the one loop correction to the fermionic $4$-point interaction involving fermions and gauge bosons to order $\Theta$ is UV finite. They have also shown that the traingle anomaly condition is hold to be true.
By calculating the one loop self-energy correction to the massless fermion to order $\Theta$, Horvat \etal (See \cite{HITY1}) show that the UV/IR 
mixing effects can be made under control. In the SW map, the first several orders of the expansion can be written in a simple form by introducing certain generalized star products \cite{JMSSW,MW}. Such an expansion enables one to treat all orders of $\Theta$ at once in each interaction vertex, thereby allows one to compute nonperturbative
results. In \cite{SY}, this technique was used to compute the fermion
one loop correction to the photon two point function of a NCQED model using SW fields. 
%For an extensive literature on the $\Theta$ resummed series, including quantum properties on the NC field theories in the Seiberg-Witten approach, see \cite{MST,HKT2,HKSTY,HITY2,HISTY1,HISTY2,HT2,Trampetic,HIKTY2}. 
%
 
%For an extensive literature on the $\Theta$ resummed series, including quantum properties on the NC field theories in the Seiberg-Witten approach, see \cite{MST,HKT2,HKSTY,HITY2,HISTY1,HISTY2,HT2,Trampetic,HIKTY2}. 
%%%%%%%%%%%%%%
 Using this SW map Calmet \etal \cite{CJSWW,CW} first constructed the {\it minimal} version of the noncommutative standard model (mNCSM in brief). They derived the ${\mathcal{O}}(\Theta)$ Feynman rules of the standard model interactions and found several new interactions which are not present in the standard model. All the above analyses were limited to the leading order in $\Theta$.  Das \etal first analysed the $e^+ e^- \to \gamma,Z \to \mu^+ \mu^-$ to order $\Theta^2$ (without considering the effect of earth's rotation)\cite{AMD}.
%%%%
There exists another version: the non-minimal version of the NCSM (nmNCSM in brief) where the triple neutral gauge boson coupling arises (absent in the mNCSM) naturally in the gauge sector. This model (nmNCSM) was first formulated by  Melic \etal \cite{MKTSW,Melic}. 
Interesting phenomenological studies comprising triple gauge interaction are available in the literature \cite{BDDSTW,AJSW,DST,BLRT1,EH1}. 

The direct test of space-time noncommutativity comes from the two decays  $Z \to \gamma \gamma,~ g g$ (forbidden in the SM at the tree level). Using the experimental bound $\Gamma^{exp}_{Z \to \gamma \gamma} < 1.3 \times 10^{-4}~\rm{GeV}$ and $\Gamma^{exp}_{Z \to g g } <  1.0 \times 10^{-3}~\rm{GeV}$, Behr \etal \cite{BDDSTW} shows the bound on the NC scale $\sim 1~\rm{TeV}$.  Taking the SM fields in the enveloping algebra, Calmet \etal \cite{Calmet1, Calmet2} shows that the bound on the NC scale $\sim 10~\rm{TeV}$, a rather weak bound. 
 
The impact of neutrino-photon interaction in the noncommutative space-time on the cooling of stars \cite{STWR}, on the primordial nucleosynthesis and ultra-high energy cosmic ray \cite{HT,HKT1} have been studied in detail. Assuming the plasmon decay to a pair of neutrino may contribute substantially to the star energy loss, the author P.~Schupp \etal \cite{STWR} obtain a lower bound $\Lambda > 81~{\rm GeV}$. The non-observation of large neutrino-nucleon cross-section for ultra high energy neutrinos (of energy $10^{10}~\rm{GeV}$ at neutrino observatories gives rise a lower bound on $\Lambda$ which is as high as $900~\rm{TeV}$\cite{HKT1}. 
The invisible $Z$ boson decay in covariant theta-exact NCSM was studied in \cite{HIKTY}, The impact of NC space-time on Quarkonia decay into two photons \cite{BMelic,TT} and $K \to \pi \gamma$ \cite{MKT} decay (forbidden in the SM) have been investigated in detail. However, the experimental upper bound on such rare decays are too weak to obtain any bound on $\Lambda$.   
Early collider searches of space-time noncommutativity include the work by Kamoshita \etal \cite{Kamoshita,Haghighat2}.  Das \etal (one of the current authors) studied in detail the Bhabha and the M\"{o}ller scattering \cite{DDR,GSDDR}, muon pair production \cite{Abhishodh1,Abhishodh2} in the non-minimal NCSM scenario. The non-commutative parameter $\Theta_{\mu\nu}$ can be fundamental constant in nature that has a fixed direction in the celestial sphere. Hence, the daily modulation effect of earth's rotation can be observed in the noncommutative phenomenology.
Now after the finding of $\sim125$ GeV scalar boson at the Large Hadron Collider, CERN in 2012, 7 TeV and 8 TeV LHC data has also established that this scalar is most likely be SM Higgs which is the last peace of SM to be discovered. Now the most important questions are as follows: whether there is any new physics beyond the standard model at the multi TeV scale which can show up in present and future colliders? And, whether the Higgs boson can throw some light in this? At the TeV energy scale where the space-time becomes noncommutative(NC), whether such a NC space-time can play an important role in Higgs boson production and it's decay?  In an earlier work one of the current authors studied the Higgs boson pair production in the nmNCSM scenario \cite{DasAbhi1,DasAbhi2}.  More recently, Wang \etal \cite{Wang} studied the Higgs-strahlung process in the context of noncommutative standard model at linear collider without considering the effect of earth's rotation into account. 
Besides the Moyal-Weyl product and Seiberg-Witten map, there is another way to deal with space-time noncommutativity. 
In this approach, one don't use any deformed products like the Moyal product: one map the field theory model onto a kind of matrix
model by representing the associative algebra modeling the
noncommutative space as an algebra of operator acting on a suitable
Hilbert space. 
%While  doing so, one is permitted to bypass the use of the Seiberg-Witten map when dealing with Moyal product.

 In this paper, we explore the associated Higgs (of mass $125~\rm{GeV}$) production with $Z$ boson at the TeV energy Linear Collider in the context of non-minimal standard model. We studied the time-averaged cross-section, azimuthal distribution, NC correction to the cross-section, the rapidity distribution (of the final state $Z$ and $H$ boson) in detail considering the effect of earth rotation into account 
%\pk{(not taken into account in \cite{Wang}) $=>$ this part not needed till referee asks specifically} 
and explore the important role played by the orientation of the NC vector ($\vec{\Theta}$) on different observables which can be tested in the upcoming TeV energy linear collider. We organise the content as follows. In Sec.~\ref{sec:process}, we discuss the noncommutative vertices involved in the process and obtain the squared matrix amplitude for the Higgs-strahlung process $ e^- e^+ \stackrel{Z^*}{\longrightarrow} Z H $ and obtained the expression of  cross-section and angular distribution with and 
without considering the effect of earth rotation into our analysis.
 
Sec.~\ref{sec:result} is devoted for the numerical analysis. We investigate the impact of space-time noncommutativity on cross-section, angular distribution, rapidity distribution. We discuss about the potential relevance of the TeV scale noncommutative geometry at the  linear collider. Finally, we summarise our results and conclude in Sec.~\ref{sec:conclusion}.

%%%%%%%%%%%%%%%%%%%%%%%%%%%%%%%%%%%%%%%%%%
\section{$ e^{-}e^{+} \rightarrow Z H $ process in nmNCSM }
\label{sec:process}
Associated Higgs production with $Z$ boson and the vector boson fusion  through $W$ or $Z$ bosons are two most significant Higgs production channels in linear collider. Both of these processes involve the VVH ($V = W^{\pm},~Z$) couplings which relates the SM symmetry breaking and precise measurements and firm constrains on anomalous couplings can be expected from linear colliders. For our present calculation,~the tree level Feynman diagram for the Higgs-strahlung process which is a $s$-channel process is shown below in Fig.~\ref{fig:feyn}. 
%%%%%%%Figure%%%%%%%%%%%%%%%
\begin{figure}[h]
\centering
\includegraphics[width=2.35in,height=2.5in,keepaspectratio]{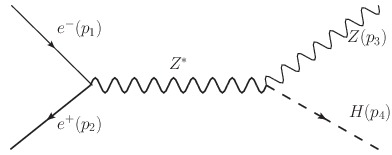}
\caption{\it Representative Feynman diagram for the process $ e^- (p_1) e^+ (p_2) \stackrel{Z^*} \to Z(p_3) H (p_4) $ both in SM and NCSM with different structure of couplings. As can be noted from the vertices given in the text, in the very large NCSM scale~($\Lambda \to \infty$) extra tensor structures disappear to reproduce SM couplings.}
\label{fig:feyn}
\end{figure}
%%%%%%%%%%%%%%%%%%%%%%%%%
We use the notation for the incoming electron and positron momenta as $p_1$ and $p_2$, whereas the outgoing $Z$ boson and Higgs($H$) boson momenta are $p_3$ and $p_4$, respectively. In terms of the noncommutative parameter $\Theta_{\mu\nu}$, we use the Feynman rule \cite{Wang} for $e^- - e^+ - Z$ vertex 
\begin{equation}
 \frac{-ie}{\sin 2\theta_{W}} \gamma^{\mu}\left\{ \left(\frac{-1}{2}+2\sin^{2} \theta_{W}\right)+\frac{1}{2} \gamma_{5}\right\} e^{i\left(\frac{p_{1}\varTheta p_{2}}{2}\right)},
\end{equation}
and for $ Z-Z-H $ vertex \\
\begin{equation}
\frac{iM^{2}_{Z}}{\upsilon} \left\{ 2\cos\left(\frac{p_{3}\varTheta p_{4}}{2}\right) \eta^{\mu \nu} + \left[ \frac{\cos\left(\frac{p_{3}\varTheta p_{4}}{2}\right)-1}{4 (p_{3}\varTheta p_{4})}\right]\left[ \left(\varTheta p_{4}\right)^{\mu}p^{\nu}_{3}+\left(\varTheta p_{4}\right)^{\nu}k^{\mu}\right]\right\}. 
\end{equation}
Neglecting the masses of incoming electrons and positrons, we consider the vertices above contain all orders of $ \Theta \left( \propto \frac{1}{\Lambda^{2}} \right)$ terms. Using these Feynman rules, we find the squared amplitude (spin-averaged) as,
%\pk{} :text will be red in color
\begin{equation}
 \overline{|M|}^{2}_{NCSM}  =  \overline{|M|}^{2}_{SM} \cos^{2}\left(\frac{p_{3}\Theta p_{4}}{2}\right)
 \label{m2NCSM}
\end{equation}
where the quantity  $p_{3} \Theta p_{4}$ (the argument of cosine function appeared in above) is given by  
\begin{eqnarray}
p_{3} \Theta p_{4} = \frac{1}{4\Lambda^{2}} \sqrt{\frac{\lambda\left(s,M^{2}_{Z},M^{2}_{H}\right)}{3}} ~f(\theta,\phi) \simeq \frac{1}{4\sqrt{3}} \left(\frac{\sqrt{s}}{\Lambda}\right)^2 f(\theta, \phi),~\rm{as}~ \sqrt{s} \gg M_Z, M_H
\label{p3Thp4}
\end{eqnarray}
where $f(\theta,\phi) = cos \theta + sin \theta \left(sin \phi + cos \phi \right) $. 
%\pk{Qn. : Is some part of oscillating behavior coming from $\cos^2{\frac{\sqrt{s}^2}{2*4*\sqrt{3}*\Lambda^{2}}}$?  Why roughly the first min coming at $\sqrt{s}^2=4*\sqrt{3}*\Lambda^{2}$ in figure (4) or (5)? What about the 2nd minimum point?}
Here $\theta$,~$\phi$ are the polar and azimuthal angles of the outgoing $Z$ boson. $\lambda\left(s,M^{2}_{Z},M^{2}_{H}\right)$ (the Kallen function), in terms of Higgs mass ($M_H$) and $Z$ boson mass ($M_Z$) are given by 
\begin{eqnarray*}
\lambda(s,M^{2}_{Z},M^{2}_{H}) = s^2 + M^{4}_{Z}+ M^{4}_{H} - 2 s M^{2}_{Z} - 2 s M^{2}_{H} - 2 M^{2}_{Z} M^{2}_{H} ~~\to s^2 ~~\rm{as}~~ \sqrt{s} \gg M_Z, M_H
\end{eqnarray*}
It is important to note that the oscillatory behaviour that the cross section and other angular dependence of the azimuthal distribution (as we will see later) is due to the presence of the function $f(\theta,\phi)$ defined above. \\ 
Note that the SM squared amplitude term gets recovered in  $ \lim_{ \Lambda \rightarrow \infty} \overline{|M|^{2}}_{NCSM}$ and is equal to 
\begin{equation}
\overline{|M|^{2}}_{SM} = \left( \frac{16 \pi \alpha M^{4}_{Z} \left[\frac{1}{4}+\left(-\frac{1}{2}+2\sin^{2} \theta_{W}\right)^{2}\right]}{\left[\left(s-M_{Z}^{2}\right)^{2}+M^{2}_{Z}\Gamma^{2}_{Z}\right]\upsilon^{2} \sin^{2}2\theta_{W}}\right)
.\left\{ \frac{s}{2}+\frac{s}{2 M^{2}_{Z}}\left(M^{2}_{Z} + \frac{\lambda\left(s,M^{2}_{Z},M^{2}_{H}\right)}{4s}\right)\right\}  
\end{equation}
%%%%%%%%%%%%%%%%%%%
Since the noncommutative parameter $\Theta_{\mu\nu}$ is considered as fundamental constant in nature, it's direction is fixed with respect to an inertial (non rotating) coordinate system. Now the experiment is done in the laboratory coordinate system which is located on the surface of the earth and is moving by the earth's rotation. As a result $\Theta_{\mu\nu}~(\vec{\Theta}_E,~\vec{\Theta}_B)$, fixed in the primary co-ordinate system, will also vary with time in the laboratory frame and this must be taken into account while making any serious phenomenological investigations of space-time.  

 To study the effect of earth's rotation  in the non-commutative space-time, we follow the notation by Kamoshita \cite{Kamoshita}.  
%{\epsfxsize=6cm\epsfbox{primary.eps}} \hspace{0.5in} {\epsfxsize=6cm\epsfbox%{primary_lab.eps}} 
%----------------------------------------------------------------------------------------------------------------------------------------------------
%%%%%%%Figure%%%%%%%%%%%%%%%
\begin{figure}[htbp]
\begin{center}
\includegraphics[width=6cm]{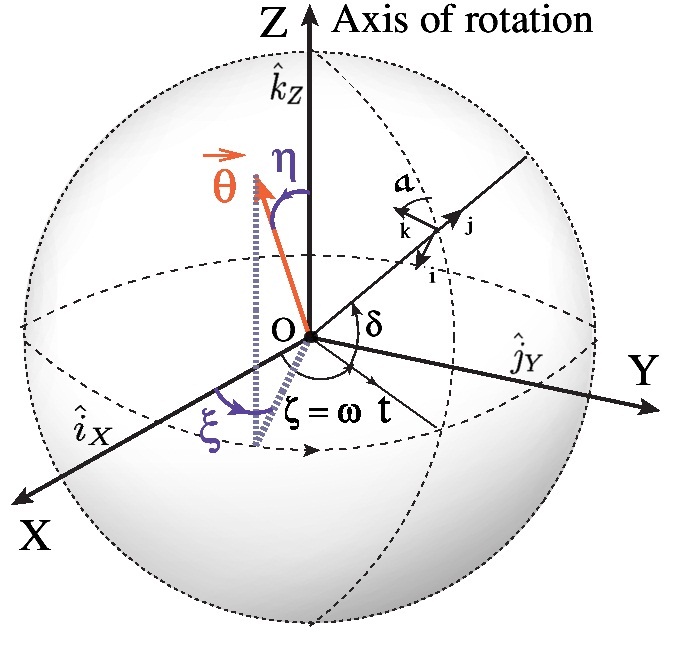}
\end{center}
\caption{{\it In the figure the primary coordinate system (X-Y-Z) with the Z-axis along the earth axis of rotation and the laboratory coordinate system (${\hat{i}}-{\hat{j}}-{\hat{k}}$) for an experiment on the earth are shown. The generic NC vector ${\vec{\Theta}}$ is shown in the X-Y-Z system where $\eta$ and $\xi$ corresponds to the polar and the azimuthal angle. In the above time varying $\zeta = \omega t$ where $\omega$ is a constant. ($\delta$,~$a$) defines the location of the laboratory.}}
\protect\label{earthplot}
\end{figure}
%---------------------------------------------------------------------------------------------------------------------
 Let ${\hat{i}}_X,~{\hat{j}}_Y$ and ${\hat{k}}_Z$ be the orthonormal basis of the primary(non rotating) coordinate system (X-Y-Z). In the laboratory coordinate system ($x-y-z$), the bases vectors are ${\hat{i}}, {\hat{j}}$ and ${\hat{k}}$. The base vectors of the primary(non rotating) coordinate system can be written in terms of the laboratory coordinate system as \cite{Kamoshita} 
\begin{eqnarray}
{\hat{i}}_X &=& (c_a s_\zeta + s_\delta s_a c_\zeta )~ \hat{i} +  c_\delta c_\zeta ~\hat{j}  + (s_a s_\zeta - s_\delta c_a c_\zeta)~ \hat{k}, \nonumber  \\
{\hat{j}}_X &=& (-c_a c_\zeta + s_\delta s_a s_\zeta)~ \hat{i} +  c_\delta s_\zeta ~\hat{j}  + (-s_a c_\zeta - s_\delta c_a s_\zeta)~ \hat{k}, \nonumber \\
{\hat{k}}_X &=&  -c_\delta s_a~  \hat{i} +  s_\delta~  \hat{j}  + c_\delta c_a~ \hat{k}
\end{eqnarray}
%%%%%%%%%%%%%%%%%%%%% 
Here $c_a = cosa,~s_a = sina,~c_\delta = cos\delta,~c_\zeta = cos\zeta$ etc. In Fig. \ref{earthplot} we have shown the primary($X - Y - Z$) and laboratory(${\hat{i}}-{\hat{j}}-{\hat{k}}$) coordinate system. 
%%Picture for the primary and laboratory coordinate system should come here ...
Note that the primary $Z$ axis lies along the axis of earth's rotation and ($\delta, a$) defines the location and orientation of $e^- - e^+$ experiment on the earth, with $- \pi/2 \le \delta \le \pi/2$ and $0 \le a \le 2 \pi$.  Because of earth's rotation the angle $\zeta$ (see Fig. 2) increases with time and the  detector comes to its original position after a cycle of one complete day, one can define $\zeta = \omega t$ with angular velocity $\omega = 2 \pi/T_{day}$ and $T_{day} = 23h56m4.09053s$. In Fig. \ref{earthplot}, the orientation of the NC vector $\vec{\Theta}$ is described by the angles $(\eta, \xi)$ in the primary coordinate system with $0 \le \eta \le \pi$ and $0 \le \xi \le 2 \pi$.
The differential cross-section of the $ e^+ e^- \stackrel{Z^*}{\longrightarrow} Z H $ scattering is given by 
\begin{eqnarray}
\frac{d\sigma}{d\cos \theta d \phi} = \frac{\lambda^{1/2} \left(s,M^{2}_{Z},M^{2}_{H}\right)}{64 \pi^{2} s^{2}}{\overline{|M|^2}_{NCSM}} 
\end{eqnarray}
and the total cross-section
\begin{eqnarray}
\sigma = \int_{0}^{\pi} d\theta \int_{0}^{2 \pi} d\phi \frac{d\sigma}{d \cos \theta~ d \phi} 
\end{eqnarray}
In above, the spin-averaged squared-amplitude  ${\overline{|M|^2}_{NCSM}} $ is given by Eqn.\ref{m2NCSM}
%%%%%%%%%%%%%%%%%%
%\beq \label{Ampsq}
%\overline {|{\mathcal{A}}|^2} = \frac{1}{4} \sum_{spin} |{\mathcal{A}}|^2 = \frac{1}{4} \sum_{spin} \left[|\mathcal{A}_\gamma|^2 + |\mathcal{A}_Z|^2 + 2 Re({\mathcal{A}}_Z {\mathcal{A}}^{\dagger}_{\gamma}) \right]
%\eeq
%{\pk Since it is difficult to get the time-dependent data}
To extract the effect of this lab-rotation coming through noncommutative effect, one takes the  average of the cross-section $\sigma$ or it's distributions over the sidereal day $T_{day}$. Here, we consider $\left<\sigma\right>_T = \frac{1}{T_{day}} \int_{0}^{T_{day}} \sigma~ dt$ and so on \cite{Kamoshita,Abhishodh2,DasAbhi2}.  
%\pk{compare that with the experimental data $=>$ Drop this part -- there is no exp data!}. 
%\bea 
%\label{sigma}
%\sigma &=& \int_{-1}^1 d(\cos\theta) \int_0^{2 \pi} d\phi \frac{d \sigma}{d\cos\theta~d\phi}, \\
%\label{dsdcostheta}
%\frac{d\sigma}{d\cos\theta} &=& \int^{2 \pi}_0 d\phi \frac{d \sigma}{d\cos\theta~d\phi},  \\
%\label{dsdphi}
%\frac{d\sigma}{d\phi} &=& \int^1_{-1} d(\cos\theta) \frac{d \sigma}{d\cos\theta~d\phi}. 
%\eea
Here $\sigma$ = $\sigma(\sqrt{s}, \Lambda, \eta,\xi, t)$. 
The cross-section is calculated using the center of mass frame of the $ e^+ e^- \stackrel{Z^*}{\longrightarrow} Z H $ process in which 4 momenta of the incoming and outgoing particles are given by:
\begin{eqnarray*}
 p_{1} & = & p_{e^-} =  \frac{\sqrt{s}}{2} \left(\ 1,0,0,1 \right), ~   p_{2} =  p_{e^+} = \frac{\sqrt{s}}{2} \left(\ 1,0,0,-1 \right), \\
 p_{3} & = &  p_{Z} = \left(\ E_{Z},~k^\prime \sin\theta ~\cos\phi,~k^\prime \sin\theta ~\sin\phi,~k^\prime \cos\theta \right), \\
 p_{4} & = &  p_{H}  = \left(\ E_{H},~-k^\prime \sin\theta~\cos\phi,~-k^\prime \sin\theta ~\sin\phi,~-k^\prime \cos\theta \right). \\
\end{eqnarray*}
where $k^\prime = \frac{1}{2}\sqrt{\frac{\lambda}{s}} $ and $\theta$ is the scattering angle made by the $3$-momentum vector $p_3$ of $Z$ boson with the $\hat{k}$ axis (the $3$-momentum direction of the incoming electron $e^-$) and $\phi$ is the azimuthal angle. The time dependence in the cross-section or it's distribution enters through the NC parameter ${\vec{\Theta}}(={\vec{\Theta}}_E)$ which changes with the change in $\zeta = \omega t $. The angle parameter $\xi$ appears in the expression of $\vec{\Theta}$ through $\cos(\omega t - \xi)$ or $\sin(\omega t - \xi)$ as the initial phase for time evolution gets disappeared in the time averaged observables. So one can deduce ${\vec {\Theta}} $ i.e. the NC scale $\Lambda$ and the orientation angle $\eta$ from the time-averaged observables.     
%%%%%%%%%%%%%%%%%%%%%
%Finally, the differential and the total cross-section for the two-body process are given by
%\begin{eqnarray}
%\frac{d\sigma}{d\Omega} = \frac{d\sigma}{d(\cos \theta) d \phi} = \frac{\lambda^{1/2} \left(s,M^{2}_{Z},M^{2}_{H}\right)}{64 \pi^{2} s^{2}}\overline{|M|^{2}} 
%\end{eqnarray}
%and 
%\begin{eqnarray}
%\sigma = \int_{0}^{\pi} d\theta \int_{0}^{2 \pi} d\phi \frac{d\sigma}{d(\cos \theta) d \phi} 
%\end{eqnarray}

%%%%%%%%%%%%%%%%%%%%%%%%%%%%%%%%
%%%%  Numerical Analysis
%%%%%%%%%%%%%%%%%%%%%%%%%%%%%%%%
%\section{Results and Discussions} % Numerical analysis
\section{Noncommutative effects on Cross-section and angular distributions}
\label{sec:result}
\begin{figure}[t]
\centering
\includegraphics[width=2.35in,height=2.5in,keepaspectratio]{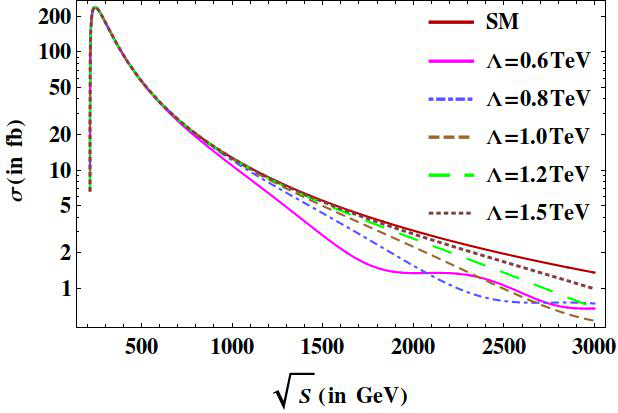}
\caption{ \it The total cross-section $\sigma$ (in fb) for associated production of Higgs with $Z$ boson is shown as a function of the machine energy $\sqrt{s}$ (in GeV). Different lines represent the choice of different non-commutative scale $\Lambda$ which is ranging from $0.6~\rm{TeV}$ to $1.5~\rm{TeV}$. The topmost curve (solid line) corresponds to the expected Standard Model production cross-section which is essentially NC cross-section at the limit $\Lambda \to \infty$.}
\label{fig:sigma_energy}
\end{figure}

We analyse the Higgs-strahlung process,  $e^-(p_1) ~ e^+(p_2) \to Z(p_3) ~ H(p_4)$ in presence of the non-commutative standard model in TeV energy linear colliders. We assume both the final $Z$ boson and Higgs boson are produced on-shell and reconstructed from their decay products which is not problematic in a linear collider by choosing suitable decay channels. The initial unpolarised electron and positron beams are colliding back to back with half of machine energy without considering any effects from ISR whereas the final particle momenta can be defined in terms of polar ($\theta$) and azimuthal angle ($\phi$) as they are produced inside the collider.
We further study this process with and without taking into consideration the effect of earth rotation into noncommutative space-time. In this context, we probe the non-commutative scale $\Lambda$ for the machine energy ranging from $500~\rm{GeV}$ to $3000~\rm{GeV}$ from the associated production of Higgs with $Z$ boson. Here we utilise the total cross-section rate, azimuthal distribution  and rapidity distributions and their time-averaged estimates to discriminate the new physics. In our analysis we have used the Higgs boson mass as $125~\rm{GeV}$ and we set the laboratory coordinate system by taking $(\delta,a) = (\pi/4, \pi/4)$ which is the OPAL experiment at LEP. 
%We also investigate the effect of the spacetime noncommutativity on the location of the earth based detector by varying  $\delta$ and $a$.  
%%%%%%%%%%%%%%%%%%%%%%%%%%%%%%
%\subsection{Cross section and angular distribution without considering the effect of earth rotation}
In Fig.~\ref{fig:sigma_energy} we display the total cross-section as a function of the center of mass energy energy $\sqrt{s}$ for different values of non-commutative scale $ \Lambda$. 
%%%%%%%%%%%%%%%%%%%%%%%%%%%%
%%%%%%%%%%%%%%%%%%%%%%%%%%%%%%%%%%%%
The uppermost curve in this plot refers the expected contribution from SM production (which corresponds to the NCSM value at the limit of NC scale $\Lambda \to \infty$ as we already noted down in previous section). While going below in the same plot, the curve next to the topmost one corresponds to the scale $\Lambda=1.5~\rm{TeV}$ and so on. On the other hand, the lowermost curve corresponds to $\Lambda = 0.6~\rm{TeV}$. 
The deviation from the SM plot starts getting manifested at and above $\sqrt{s} = 1~\rm{TeV}$. For example, at the machine energy $\sqrt{s} = 1.5~\rm{TeV}$, we see that the overall cross-section gradually increases with the increase in the scale $\Lambda$ from $0.6~\rm{TeV}$ to $1.5~\rm{TeV}$ before merging with the SM value for larger NC scale. Expectedly, maximum deviation from that of SM in observed for the lower  values of  $\Lambda$. Also, note that for a given machine energy the NC cross-section is always less than that the corresponding SM value. That is simply followed from the Eq.~\ref{m2NCSM}. Moreover, one can notice that at higher machine energy the NC cross-sections are not always simply falling with monotonous regularity. The ripple effect appears and become prominent at some higher machine energy subject to each NC curves. This is an effect of tensor structure coming into the Eq.~\ref{m2NCSM}. In the next subsection we would explore this as a characteristic feature from NC effects. \\\\

%%%%%%%%%%%%%%%%%%%%%%%%%%%%%%%%%%%%
%\noindent {\bf {Noncommutative correction}} \\
\subsection{Noncommutative correction}
To explore and quantify the noncommutative effect manifested in production cross-section and originated from the tensorial strunture as in $ \Theta_{\mu\nu}$, we define the NC correction with respect to the SM value $\Delta \sigma$ as, \\ 
 \begin{equation}
 \Delta \sigma = \sigma_{NC} - \sigma_{SM}.
 \end{equation}
%%%%%%%%%%%%%%%%%%%%%%%%%%%%%%%%%%%%%%%%%%%%%%%%%%%%
\begin{figure}[t]
\centering
\includegraphics[width=2.35in,height=2.5in,keepaspectratio]{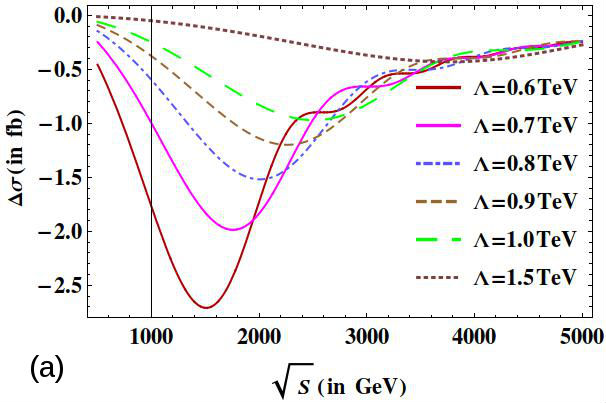} 
\includegraphics[width=2.35in,height=2.5in,keepaspectratio]{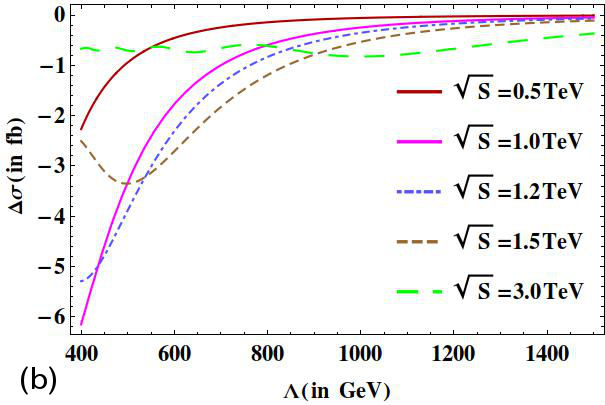}
\caption{\it On the left, (a) noncommutative correction to the total cross-section $\Delta \sigma$ (in fb) for associated production of Higgs with $Z$ boson is shown as a function of the linear collider machine energy $\sqrt{s}$ (in GeV) for different $\Lambda$ ranging from $0.6~\rm{TeV}$ to $1.5~\rm{TeV}$. On the right, (b) the same quantity is plotted as a function of  NC scale $\Lambda$ for different machine energy.}
\label{fig:del_sigma_vs_rts}
\end{figure}
%%%%%%%%%%%%%%%%%%%%%%%%%%%%%%%%%%%%%%%%%%%%%%%%%%%%%
In Fig.~\ref{fig:del_sigma_vs_rts}(a), we have demonstrated the variation of $\Delta \sigma$ as a function of the machine energy $\sqrt{s}$ for different values of $\Lambda$. Now, we can spot it very clearly that $ \Delta \sigma $  first decreases (becomes more negative as $\sigma_{NC} < \sigma_{SM}$) with the increase in $\sqrt{s}$ and reaches a minimum for its first trough \footnote{However additional subsequent troughs remain subdominant to be observed in this figure and would be clearly prominent in our next figure where we would consider the relative correction}. Note that the first minimum occurs at $\sqrt{s} = 2\times(3)^{1/4}\times\Lambda$ (See Eqn.\ref{p3Thp4})$= 1579(2105)~\rm{GeV}$ corresponding to $\Lambda= 600(800)~\rm{GeV}$,whereas the second minimum occurs at $\sqrt{s} \approx 2700(3600)~\rm{GeV}$
for the same $\Lambda$.  After that they decreases with the increase in $\sqrt{s}$ and pass through an oscillatory phase and eventually becomes a flat curve asymptotically meeting at the line $\Delta \sigma=0$, which is the case for large $\Lambda$ limit(the SM limit).
%The oscillatory behaviour is due to the presence of the  term $ Cos\left\{ \left(\frac{\sqrt{S}}{\Lambda} \right)^{2} f(\theta,\phi) \right\}$ in the NC cross-section. 
As for example, for $\Lambda = 0.6~\rm{TeV}$, we find $\Delta \sigma = -2.7~\rm{fb}$ at $\sqrt{s} = 1.5~\rm{TeV}$ and for $\Lambda = 0.6~\rm{TeV}$, $\Delta \sigma = -0.4~\rm{fb}$ (minimum) at $\sqrt{s} = 4.0~\rm{TeV}$. As we vary $\Lambda = 0.6~\rm{TeV}$ to $\Lambda = 1.5~\rm{TeV}$, we see that the height/depth of the trough $\Delta \sigma_{min}$ decreases and it's location (the value of $\sqrt{s}$) also gets changed. In Fig.~\ref{fig:del_sigma_vs_rts}(b), we have plotted the variation of the difference $\Delta \sigma$ as a function of the NC scale $\Lambda$ corresponding to the different machine energy $\sqrt{s} = 0.5,~1.0,~1.2,~1.5$ and $3~\rm{TeV}$, respectively. We see that the deviation decreases with the increase in $\Lambda$ for a fixed machine energy and for large $\Lambda$ the NC effect gets disappeared as expected. For $\sqrt{s} = 1 ~\rm{TeV}$, $|\Delta \sigma|$ changes by an amount $6~\rm{fb}$ as $\Lambda$ changes from $400~\rm{GeV}$ to $1500~\rm{GeV}$.  

As noted earlier, we next define the relative correction of this NC cross-section by normalising the noncommutative correction to the total cross-section by the SM value: 
 \begin{equation}
 \delta_{r} = \frac{\Delta \sigma }{\sigma_{SM}}.
 \end{equation} \\
%%%%%%%%%%%%%%%%%%%%%%%%%%%%%%%%%%%%%%%% 
 \begin{figure}[h]
\centering
\includegraphics[width=2.35in,height=2.5in,keepaspectratio]{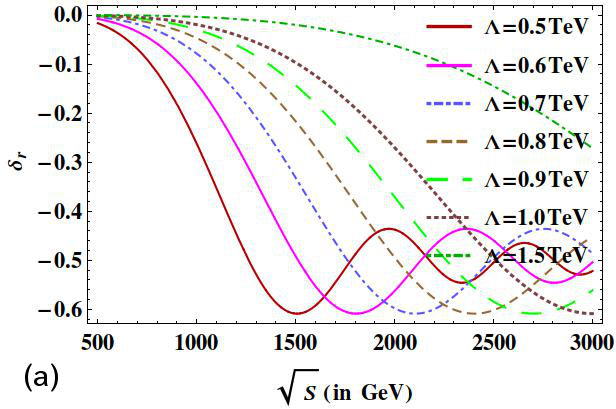}
\includegraphics[width=2.35in,height=2.5in,keepaspectratio]{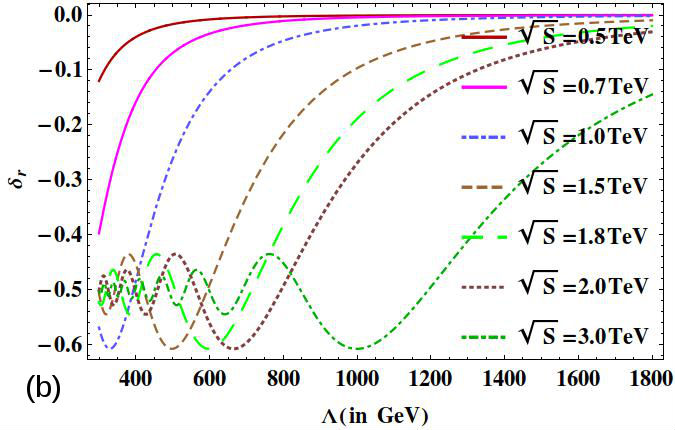}
\caption{{ \it On the left, (a) the ratio $\delta_r$ is plotted as a function of $\sqrt{s}$ (in GeV) for different $\Lambda$ values. On the right, (b) $\delta_r$ is shown as a function of $\Lambda$ (in TeV) for different machine energy.}}
\label{fig:del_sigma_by_sigma_vs_rts}
\end{figure}
%%%%%%%%%%%%%%%%%%%%%%%%%%%%%%%%%%%%%%%%
In Fig.~\ref{fig:del_sigma_by_sigma_vs_rts}, we have shown the variation of $\delta_r$ against the machine energy $\sqrt{s}$ for different values of $\Lambda$ (left figure), whereas on the right figure, we have shown  the dependence of $\delta_r$ against the NC scale $\Lambda$  for few different machine energy $\sqrt{s}$. From the left figure, we see that $\delta_r$ becomes maximum at its first peak when $\sqrt{s} = 1.5,~1.8,~2.1,~2.4~,2.7~\rm{TeV}$ and $3.0~\rm{TeV}$ corresponding to  $\Lambda = 0.5,~0.6,~0.7,~0.8,~0.9~\rm{TeV}$ and $1.0~\rm{TeV}$. For $\Lambda = 0.5$ TeV at $ \sqrt{s} = 1.5$ TeV, the deviation $\delta_{r}$ reaches it's minimum value $ -0.6085$. This is called primary minimum (crest). Similarly the secondary minimum (crest) corresponds to $ \delta_{r} = -0.5456 $ and  the next to it corresponds to $ \delta_{r} = -0.435$. 
%Finally at large $\sqrt{s}$, $ \delta_{r}$ converge to $ -0.5$ for all $\Lambda$. 
On the right plot, we have shown how $\delta_r$ varies with $\Lambda$ as the machine energy increases from $0.5~\rm{TeV}$ to $3~\rm{TeV}$. For $\Lambda \le 1~\rm{TeV}$, we see that for a wide range of machine energies, $\delta_r$ converges to $-0.5$. In Table ~\ref{tab:correction}, we have displayed the value $\delta_r$ for different $\Lambda$ corresponding to different machine energies. We also listed several values of $\delta_r$ for different $\Lambda$ corresponding to CLIC energy ($\sqrt{s} = 3.0~\rm{TeV}$).  We find that $\delta_r(=-0.608596)$ is maximum corresponding to $\Lambda \sim 1~\rm{TeV}$.
%%%%%%%%%%%%%%%%%%%%%%%%%%%%%%%%%%%%%%%%%
\begin{table*}[t]
\caption{ The NC correction $ \delta_{r}$ against the NC scale $\Lambda$ (in GeV) is shown corresponding to different machine energy $\sqrt{s}$. Primary peak values and corresponding NC scales are shown in bold. }
\begin{center}
\begin{tabular}{|c|c|c|c|}
 \hline 
 $ \sqrt{s} = 500$~GeV & $ \sqrt{s} = 1$~TeV &  $ \sqrt{s} = 1.5$~TeV  & $ \sqrt{s} = 3$~TeV  \\ \hline \hline
 \begin{tabular}{c|c}         
   $ \Lambda$ (GeV) & $ \delta_{r}$ \\  \hline \\
   100  & -0.543 \\
   {\bf 158} & {\bf -0.608} \\
   160 & -0.607 \\
   250 & -0.229 \\
   300 & -0.120 \\
   600 & -0.008 
 \end{tabular}
 & \begin{tabular}{c|c}
 $ \Lambda $ (GeV) & $ \delta_{r}$ \\ \hline \\
   200  & -0.515 \\
   300 & -0.569 \\
   {\bf 330} & {\bf -0.608} \\
   400 & -0.485 \\
   600 & -0.139\\
           &   \\
 \end{tabular} 
 & \begin{tabular}{c|c}
 $ \Lambda$ (GeV) & $ \delta_{r}$ \\ \hline \\
   300 & -0.508 \\
   450 & -0.564 \\
   {\bf 497} & {\bf -0.608} \\
   600 & -0.492 \\
       &         \\
       &
 \end{tabular} 
 & \begin{tabular}{c|c}
 $ \Lambda$ (GeV) & $ \delta_{r}$ \\ \hline \\
   600 & -0.504 \\
   700 & -0.485 \\
   800 & -0.452 \\
   {\bf 999} & {\bf -0.608} \\
   1100 & -0.572 \\
       &         
 \end{tabular} 
 \\ \hline
 \end{tabular}
\end{center}
\label{tab:correction}
 \end{table*} 
%%%%%%%%%%%%%%%%%%%%%%%%%%%%%%%%%%%%%%
%%%%

%\noindent {\bf {Azimuthal distribution of the $Z$ and the Higgs boson in absence of earth rotation}} \\ \\
%\subsection{Azimuthal distribution of $Z$ and Higgs boson in absence of earth rotation}
\subsection{Angular distributions in absence of earth rotation}

The angular distribution of the final state scattered particles is a useful tool to understand the nature of new physics. Since the noncommutativity of space-time breaks Lorentz invariance including rotational invariance around the beam axis, this will lead to an anisotropy in the azimuthal distribution of the cross-section i.e. the distribution will depends strongly on $\phi$. 
In the standard model the azimuthal distribution for one of the final particles is found to be flat. However, in the noncommutative standard model(NCSM) due to the presence of the tensor $\theta$-weighted dot product i.e. terms like $p_3 \theta p_4 \sim \left(\frac{\sqrt{s}}{\Lambda}\right)^2 \left(cos \theta + sin \theta \left(sin \phi + cos \phi \right) \right)$ (Eqn.~\ref{p3Thp4}), these distributions are no longer remain flat.  
%%%%%%%%%%%%%%%%%%%%%%%%%%%%%%%%%%%%%%
\begin{figure}[h]
\centering
\includegraphics[width=2.0in,height=2.0in,keepaspectratio]{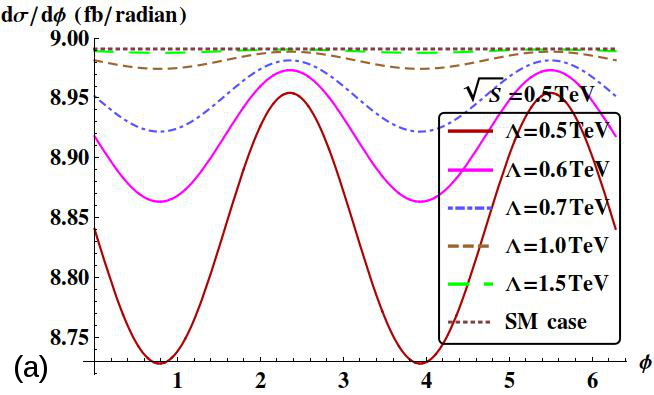}
\includegraphics[width=2.0in,height=2.0in,keepaspectratio]{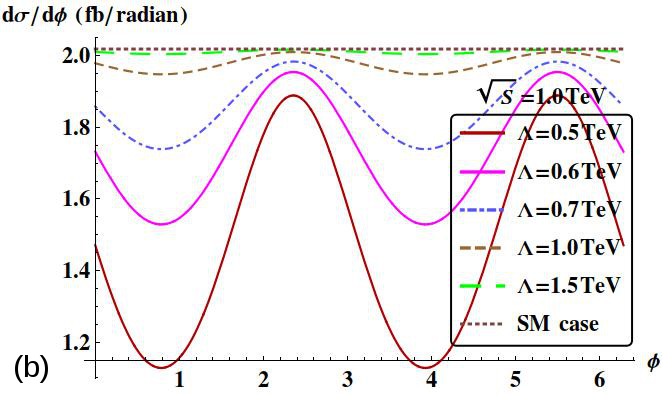}
\includegraphics[width=2.0in,height=2.0in,keepaspectratio]{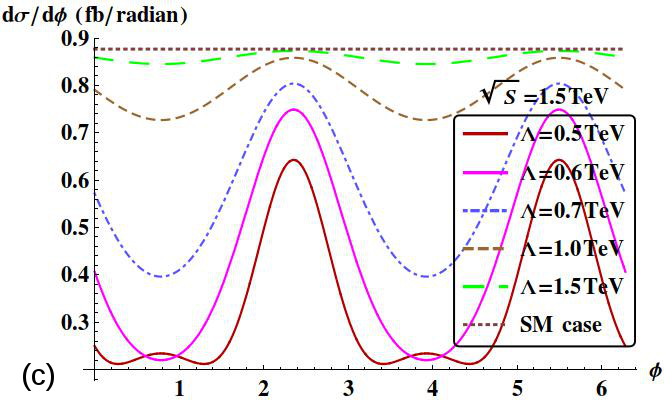}
\includegraphics[width=2.0in,height=2.0in,keepaspectratio]{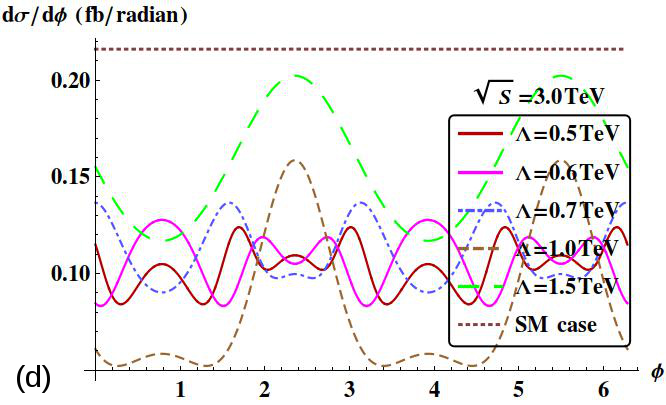}
\caption{\it The azimuthal distribition $\frac{d\sigma}{d\phi}$ (in fb/rad) is plotted as a function of $\phi$ (in rad) for different $\Lambda$ values. Displayed are four plots (a), (b) and  (c) corresponding to different machine energy $\sqrt{s} = 0.5,~1.0,~1.5$ and $3.0~\rm{TeV}$ are shown. }
\label{fig:phi}
\end{figure}
%%%%%%%%%%%%%%%%%%%%%%%%%%%%%%%%%%%%%
We see (Eqs.~\ref{m2NCSM}, \ref{p3Thp4}) that the squared amplitude is a oscillatory function of $ (\theta,\phi)$ and is further distorted by an  oscillatory function of $ (\sqrt{s},\Lambda)$ (of the form $\sim cos\left\{\left(\frac{\sqrt{s}}{\Lambda}\right)^2~\left(cos \theta + sin \theta \left(sin \phi + cos \phi \right) \right) \right\}$ ).
In Fig.~\ref{fig:phi}, we have shown the azimuthal $\phi$ distributions for different $\Lambda$ values corresponding to the machine energy $\sqrt{s} = 0.5,~1.0,~1.5$ and $3.0~\rm{TeV}$, respectively.
The distribution has several maxima(minima) located at $\phi = 2.4,~5.4~\rm{rad}$ ($0.8,~4.0~\rm{rad}$). For a given machine energy (e.g. $\sqrt{s} = 1.0~\rm{TeV}$), the height of the peaks decreases with the increase in $\Lambda$.  We see that as the machine energy is increased to $3~\rm{TeV}$, the peaks corresponding to $\Lambda = 0.5,~0.6,~0.7~\rm{TeV}$ located at $\phi = 2.4,~5.4~\rm{rad}$  gets smeared, while the peaks corresponding to $\Lambda = 1.0,~1.5~\rm{TeV}$ still  survives. The fluctuation observed at the crest and trough of the figure corresponding to machine energy $\sqrt{s} = 3~\rm{TeV}$ is due to the fact that $\frac{d \sigma}{d \phi}$ is a oscillatory function of $\phi$ and as well as the oscillatory function of $ (\sqrt{s},\Lambda)$. At higher energy $ \frac{d \sigma}{d \phi}$ is dominated by $\sqrt{s}$ and $ \Lambda$. \\

%%%%%%%%%%%%%%%%%%%%%%%%%
Next, we define the rapidity of a particle as 
\begin{equation*}
 y = \frac{1}{2} ln \left( \frac{E -  P_z}{E + P_z} \right)
\end{equation*}
where $E$, the energy and $P_z$, the $z$-component momentum of the particle ($Z$ boson or the Higgs boson $H$).
%%%%%%%%%%%%%%%%%%%%%%%%
\begin{figure}[htb]
 \centering
\includegraphics[width=1.9in,height=2.9in,keepaspectratio]{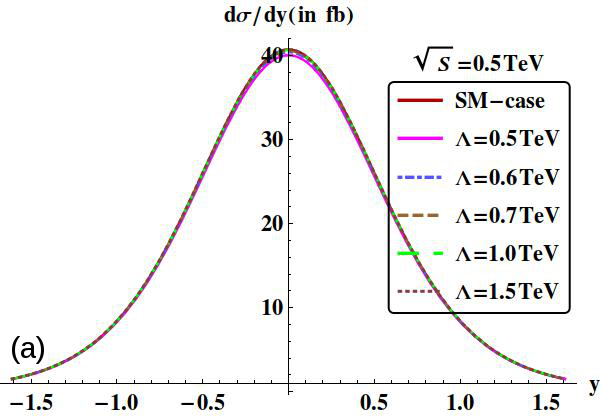}
\includegraphics[width=1.9in,height=2.9in,keepaspectratio]{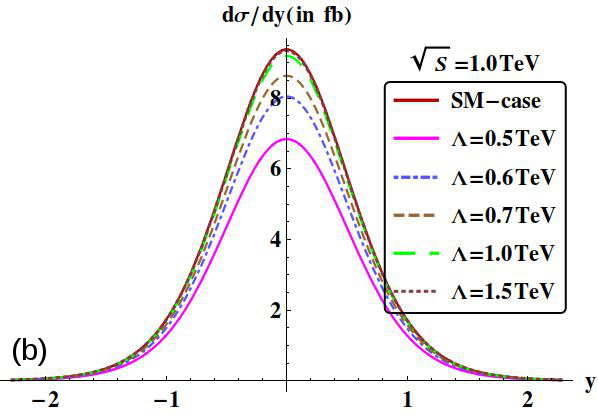}
\includegraphics[width=1.9in,height=2.9in,keepaspectratio]{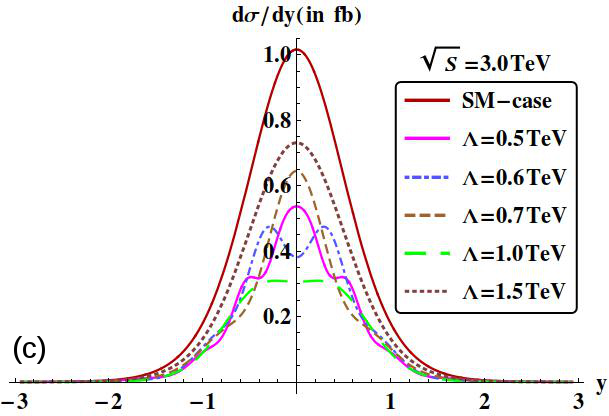}
\caption{\it The rapidity distribution $\frac{d\sigma}{dy}$ (in fb) is plotted as a function of the rapidity $y$ for different $\Lambda$ values. Three plots (a), (b) and (c) are shown corresponding to machine energy $\sqrt{s} = 0.5,~1.0$ and $3.0~\rm{TeV}$, respectively.}
\label{fig:rapidityNR}
\end{figure}
%%%%%%%%%%%%%%%%%%%%%%%
In Fig.~\ref{fig:rapidityNR}, we have plotted the distribution $d\sigma/dy$ as a function of rapidity $y$ for different cases of machine energies. In the leftmost Figure where the machine energy is fixed at $\sqrt{s}= 0.5~\rm{TeV}$, we see that the height of the peak (located at $y=0$) increases with the increase in $\Lambda$. The topmost curve corresponds to the SM curve ($\Lambda \to \infty$). As we move towards left (i.e. increase the machine energy from $0.5~\rm{TeV}$ to $3.0~\rm{TeV}$, the height of a peak (corresponding to a particular $\Lambda$ value ) decreases and gets flattened. When the machine energy is equal to $3~\rm{TeV}$, the peak at $y=0$ corresponding to $\Lambda = 0.6~\rm{TeV}$ split into two. Note that the rapidity distribution of $Z$ boson or Higgs particle are the same.\\

%%%%%%%%%%%%%%%%%%%%%%%%%%%%%%%%%%%%%%%%%%%%%%%%%%%%%%%%%%%%%%%%%%
%% On analysis Earth rotational effect is taken into account
%%%%%%%%%%%%%%%%%%%%%%%%%%%%%%%%%%%%%%%%%%%%%%%%%%%%%%%%%%%%%%%%%%
%\noindent {\bf The cross-section, angular distribution and the effect of earth rotation} \\
\section{Consequence of earth rotation on the cross-section and angular distributions}
\label{sec:rotation}
\subsection{Cross-section, diurnal motion in presence of earth rotation}
It was pointed out earlier that one significant aspect for noncommutative effect can be originated from directionality of fundamental NC fix points. The experiment is done in the laboratory frame attached to the earth surface which is rotating, whereas directions related to the NC parameter $\Theta_{\mu\nu}$ has a fixed direction in the celestial sphere. In Sec.2 we described the notation to parametrise the processes in rotating frame. This rotation can have a direct but subleading impact in daily modulation on the inherent structure of the interaction couplings. Noncommutative contributions and angular dependence due to additional tensorial were presented in Se. 3.1, where this effect was not considered. Now we would like to analyse the effect of earth rotation on the orientation of the NC vector $\vec{\Theta}_E$ and thus on the cross-section and angular distribution of the associated Higgs production. Since the cross-section and the angular distributions are function of time, we made a time-averaged (
i.e. averaged over the side-real day $T_{day}$) estimate of the total cross-section, correction, azimuthal distribution 
and rapidity distributions to account for this additional effect coming from the new physics. The laboratory coordinate system is being set at $(\delta,a) = (\pi/4, \pi/4)$ which is the OPAL experiment at LEP. Our choice of same lab-system enables one to directly compare the consequence due to the earth rotation. The time-averaged azimuthal distribution and the cross-section are defined as, 
\bea \label{dsigma_avg}
\left<\frac{d\sigma}{d\phi}\right>_T &=& \frac{1}{T_{day}} \int_{0}^{T_{day}} \frac{d\sigma}{d\phi} dt = \frac{1}{T_{day}} \int_{0}^{T_{day}} \int_{-1}^{1} \frac{d\sigma}{dcos\theta ~d\phi} dcos\theta ~dt, \\
\left<\sigma\right>_T &=& \frac{1}{T_{day}} \int_{0}^{T_{day}} \sigma dt = \frac{1}{T_{day}} \int_{0}^{T_{day}} \int_{-1}^{1} \int_{0}^{2 \pi} \frac{d\sigma}{d\phi ~dcos\theta} dcos\theta~d\phi ~dt,
\eea
where $T_{day} = 23{\rm h}56{\rm m}4.09053{\rm s}$, the sidereal day. 

\begin{figure}[htb]
 \centering
\includegraphics[width=2.25in,height=2.25in,keepaspectratio]{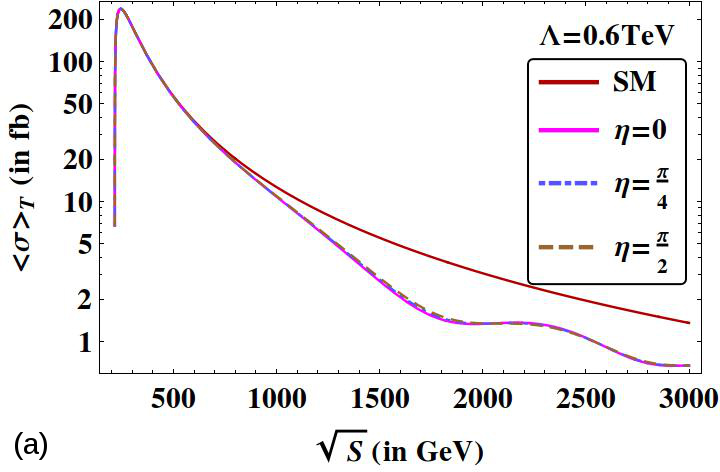}
\includegraphics[width=2.25in,height=2.25in,keepaspectratio]{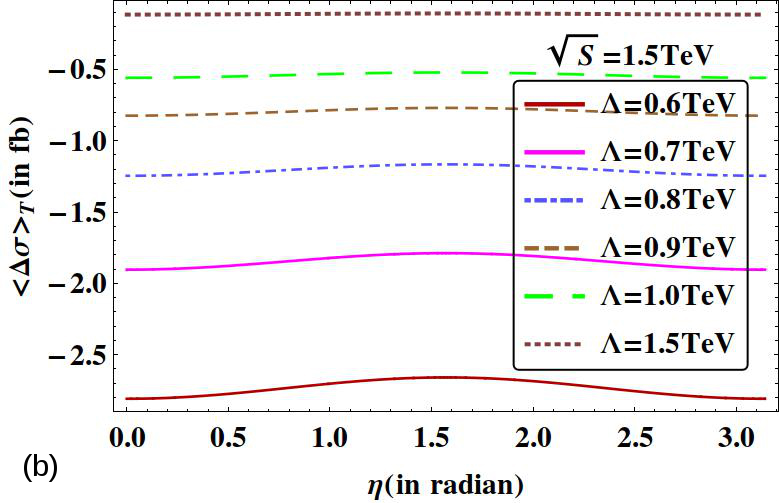}
\caption{\it (a) The time-averaged total cross-section $\langle \sigma \rangle_T$ (in fb) for associated production of Higgs with $Z$ boson is shown as a function of the linear collider center of mass energy $\sqrt{s}$. The topmost curve (solid line) corresponds to the expected Standard Model production cross-section which is essentially NC cross-section at the limit $\Lambda \to \infty$. The three other plots (below the SM plot) corresponds to $\eta = 0,~\pi/4$ and $\pi/2$ and $\Lambda = 0.6~\rm{TeV}$ are found to be almost overlapping. (b) The time-averaged NC correction to the cross-section $\langle \Delta \sigma \rangle_T$ (in fb)  is shown as a function of orientation angle $\eta$ of the NC vector for a fixed machine energy $\sqrt{s} = 1.5~\rm{TeV}$. The different plots correspond to $\Lambda = 0.6,~0.7,~0.8,~0.9,~1.0$ and $1.5~\rm{TeV}$.}
\label{fig:cs_energy_LR}
\end{figure}

%%%%%%%%%%%%%%%%%%%%%%%%%%%%%%
In Fig.~\ref{fig:cs_energy_LR} we have shown the time-averaged cross-section $\langle \sigma \rangle_T$ as a function of the machine energy $\sqrt{s}$ corresponding to $\eta = 0,~\pi/4$ and $\pi/2$.
%%%%%%%%%%%%%%%%%%%%%%%%%%%%%%% 
%\end{document}
%%%%%%%%%%%%%%%%%%%%%%%%%%%%%%%%%%
 Only one NC scale ($\Lambda = 0.6~\rm{TeV}$) is chosen for demonstration comparing with the corresponding non-rotation plot in Fig. 3. The plots corresponding to $\eta=0,~\pi$ and $\eta = \pi/2$ are seen to be nearly overlapping  with a narrow effect due to different choices of $\eta$ values. To demonstrate the variation due to this parameter, we next define the NC-correction to the cross-section (time-averaged) as $\langle \Delta \sigma \rangle_T= \langle \sigma_{NC} \rangle_T - \langle \sigma_{SM} \rangle_T $. In Fig. 8(b), we have plotted $\langle \Delta \sigma \rangle_T$ as a function of $\eta$ corresponding to $\Lambda = 0.6,~0.7,~0.8,~0.9,~1.0$ and $1.5$ TeV for a fixed machine energy $\sqrt{s} = 1.5~\rm{TeV}$. The plot shows a peak at $\eta = \pi/2$ which corresponds to the fact that $\langle \sigma_{NC} \rangle_T$ is larger at that value irrespective to the $\Lambda$ chosen. However this maximum deviation is quite small; around $0.1~{\rm fb}$ corresponding to $\Lambda = 0.6~\rm{TeV}$ and even less 
for larger $\Lambda$. 
%%%%%%%%%%%%%%%%%%%%%%%%%%%%%%%%%%%%%%%%%%%%%%%%%%%%%%%%%%%%%%%%%%%%% 
\begin{figure}[htb]
 \centering
\includegraphics[width=2.25in,height=2.25in,keepaspectratio]{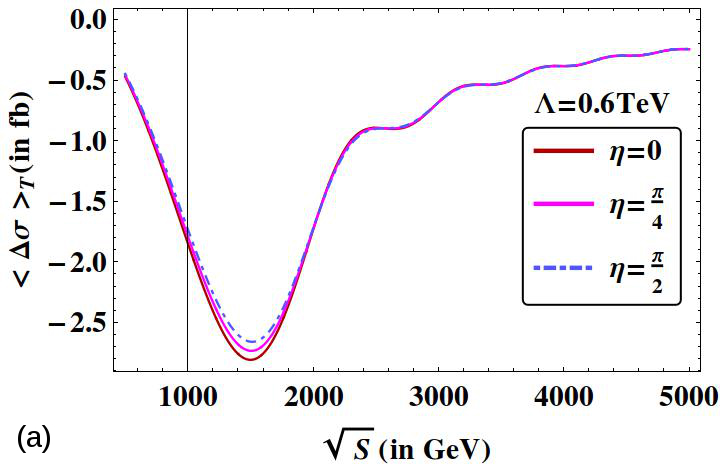}
\includegraphics[width=2.25in,height=2.25in,keepaspectratio]{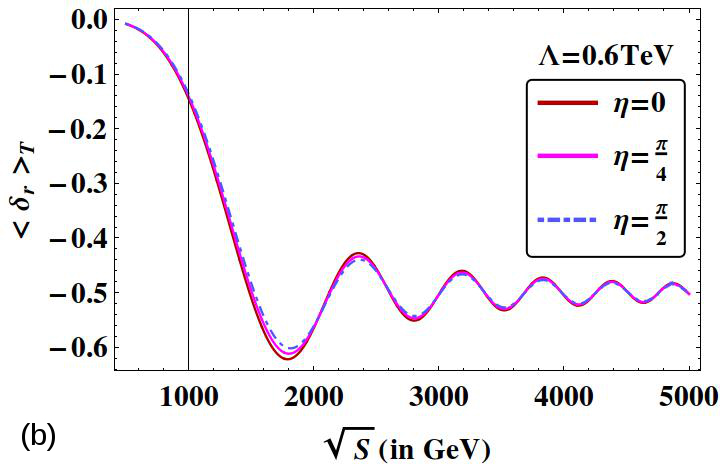}
\caption{\it In (a) the time-averaged NC correction to the cross-section $\langle \Delta \sigma \rangle_T$ (in fb)  is shown as a function of the machine energy $\sqrt{s}$ (in GeV) for $\Lambda = 0.6~\rm{TeV}$. The different plots correspond to $\eta = 0, \pi/4, \pi/2$. In (b), the normalised correction $\langle \delta_r \rangle_T$ is shown as a function of $\sqrt{s}$ for the same above set of $\eta$ values.}
\label{fig:delsigdelr_rts}
\end{figure}
%%%%%%%%%%%%%%%%%%%%%%%%%%%%%%%%%%%%%%%%%%%%%%%%%%%%%%%%%%%%%%%%%%%%%%%%

In Fig.\ref{fig:delsigdelr_rts}a and Fig.\ref{fig:delsigdelr_rts}b, we have shown  $\langle \Delta \sigma \rangle_T$ and $\langle \delta_r \rangle_T$ as a function of $\sqrt{s}$ for $\Lambda = 0.6~\rm{TeV}$ and $\eta = 0,~\pi/4$ and $\pi/2$, respectively. 
We see that the curves corresponding to different $\eta$ are almost overlapping for different machine energy values except near the region $\sqrt{s} = 1.5~\rm{TeV}$. From the lowermost to the topmost curves $\eta$ corresponds to $0,~\pi/4$ and $\pi/2$, respectively. We see that for $\eta = \pi/2$, the deviation  $\langle \Delta \sigma \rangle_T$ and the normalised deviation $\langle \delta_r \rangle_T =  {\langle \Delta \sigma \rangle_T }/{\langle \sigma_{SM} \rangle_T}$ are minimum, yielding $\langle \sigma_{NC} \rangle_T$ is largest for $\eta = \pi/2$. We set $\eta = \pi/2$ in the rest of our analysis. 
As expected from these results, one finds that the variation of time-averaged quantity $\langle \Delta \sigma \rangle_T$ and $\langle \delta_{r} \rangle_T$ with respect to machine energy $\sqrt{s}$ (for different $\Lambda$) or $\Lambda$ (for different $\sqrt{s}$) remains very similar to the plots shown in Fig~\ref{fig:del_sigma_vs_rts} and Fig.~\ref{fig:del_sigma_by_sigma_vs_rts} in absence of the earth rotation with a subleasing shift. We skip the repetition of these plots for the brevity. Magnitude of the shifts for different parameters  are better expressed in a representative table similar to the one we discussed in last section.
In Table ~\ref{tab:correction_LR}, we have displayed time-averaged $\langle \delta_r\rangle_T$ for different $\Lambda$ corresponding to different machine energies and $\eta = \pi/2$ after taking consideration of earth rotation effect. 
%%%%%%%%%%%%%%%%%%%%%%%%%%%%%%%%%%%%%%%%%%%%%%%%%%%%%%%%%%%
\begin{table*}[t]
\caption{ The NC correction $\langle \delta_r\rangle_T$ against the NC scale $\Lambda$ is shown corresponding to different machine energy $\sqrt{s}$ and orientation angle of the NC vector $\eta = \pi/2$. Primary peak value  and the corresponding NC scales are shown in bold.}
\begin{center}
\begin{tabular}{|c|c|c|c|}
 \hline 
 $ \sqrt{s} = 500$GeV & $ \sqrt{s} = 1000$GeV &  $ \sqrt{s} = 1500$GeV  & $ \sqrt{s} = 3000$GeV  \\ \hline \hline
 \begin{tabular}{c|c}         
   $ \Lambda$ GeV & $ \langle \delta_{r} \rangle_{T} $ \\  \hline \\
   100  & -0.541 \\
   {\bf 158} & {\bf -0.604} \\
   160 & -0.603 \\
   250 & -0.226 \\
   300 & -0.118 \\
   600 & -0.008
 \end{tabular}
 & \begin{tabular}{c|c}
 $ \Lambda $ GeV & $ \langle \delta_{r} \rangle_{T} $ \\ \hline \\
   200  & -0.514 \\
   300 & -0.568 \\
   {\bf 330} & {\bf -0.602} \\
   400 & -0.476 \\
   600 & -0.136
 \end{tabular} & \begin{tabular}{c|c}
 $ \Lambda$ GeV & $ \langle \delta_{r} \rangle_{T} $ \\ \hline \\
   300 & -0.508 \\
   450 & -0.563 \\
   {\bf 497} & {\bf -0.602} \\
   600 & -0.483 \\
       &         \\
       &
 \end{tabular} & \begin{tabular}{c|c}
 $ \Lambda$ GeV & $ \langle \delta_{r} \rangle_{T} $ \\ \hline \\
   600 & -0.504 \\
   700 & -0.485 \\
   800 & -0.456 \\
   {\bf 999} & {\bf -0.602} \\
   1100 & -0.563 \\
       &         \\
       & 
 \end{tabular} 
 \\ \hline
 \end{tabular}
\end{center}
\label{tab:correction_LR}
 \end{table*} 
%%%%%%%%%%%%%%%%%%%%%%%%%%%%%%%%%%%%%%%%%%%%%%%%%%%%%%%%%%%
Note, for example, the shift in the ratio $|\langle \delta_{r} \rangle_{T}| = 0.604$  form our non-rotating estimate of $|\langle \delta_{r} \rangle| = 0.608$ for values of $\Lambda = 158~\rm{GeV}$ and $\sqrt{s} = 500~\rm{GeV}$.

Diurnal modulation in the production signal in our lab frame can also appear as a distinctive feature of fundamental NC fixed points. This modulation would depend upon $\eta$ together with the NC scale $\Lambda$ and machine energy $\sqrt{s}$, although the phase of these oscillation are fixed by the choice of experiment location. 
%\pk{position of these oscillation peaks $=>$ } 

%%%%%%%%%%%%%%%%%%%%%%%%%%%%%%%%%%%%%%%%%%%%%%%%%%%%%%%%%%%%%%%%%%%%%%%%%%
\begin{figure}[htb]
 \centering
\includegraphics[width=2.0in,height=2.0in,keepaspectratio]{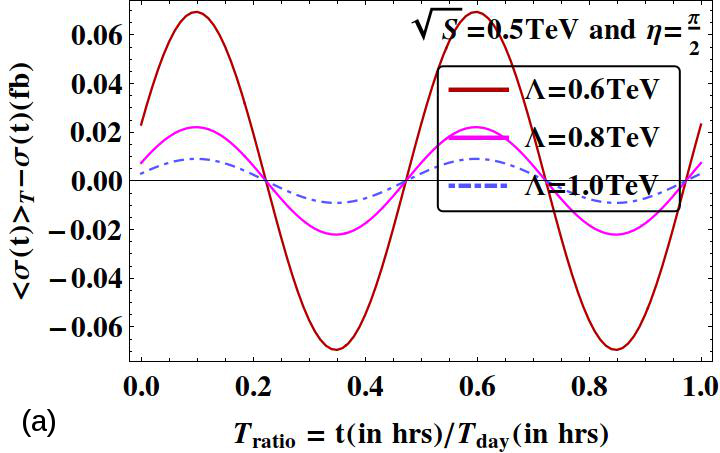}
\includegraphics[width=2.0in,height=2.0in,keepaspectratio]{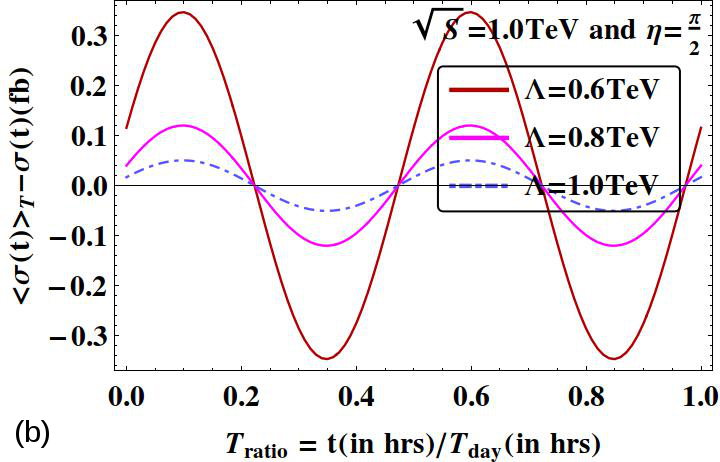}
\includegraphics[width=2.0in,height=2.0in,keepaspectratio]{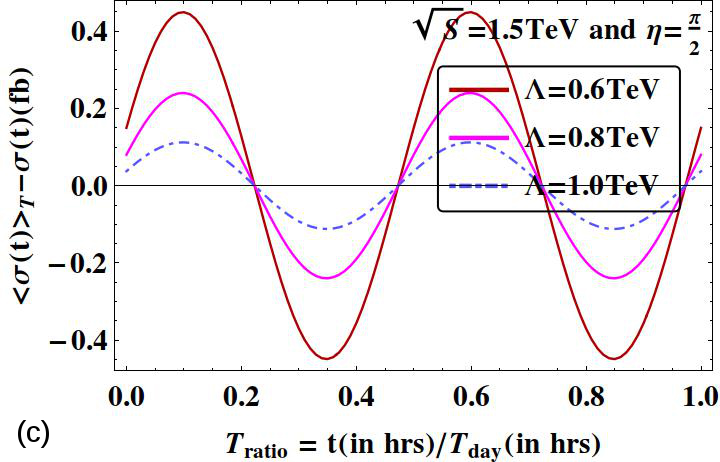}
\includegraphics[width=2.0in,height=2.0in,keepaspectratio]{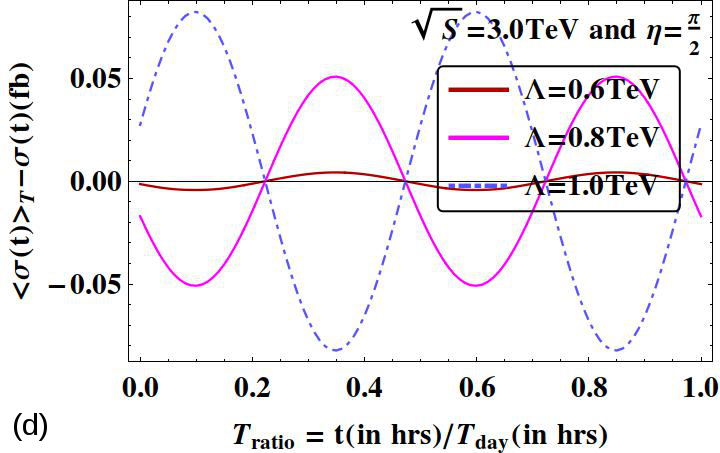}
\caption{\it The diurnal modulation $\Delta(t)$ in the production signal is plotted as a function of time fraction of sidereal day $T_{ratio}(=t/T_{day})$ for the machine energy $\sqrt{s}=0.5,~1.0,~1.5$ and $3.0~\rm{TeV}$, respectively. The NC scale is chosen as $\Lambda = 0.6,~0.8$ and $1.0$~TeV and $\eta = \pi/2$.}
\label{fig:siderealrate}
\end{figure}
%%%%%%%%%%%%%%%%%%%%%%
To take into account this modulation, we define the quantity $\Delta(t) = \sigma(t) - \langle \sigma(t) \rangle_T$. This is the deviation of the Higgs-strahlung production cross-section $\sigma(t)$ at any time from the time-averaged cross-section $\langle \sigma(t) \rangle_T$ over the period of sidereal day, $T_{day}=23.934~\rm{hours}$.  In Fig. \ref{fig:siderealrate}, we have plotted  $\Delta(t)$ as a function time fraction for the machine energy $\sqrt{s}=0.5,~1.0,~1.5$ and $3.0~\rm{TeV}$, respectively. 
In each plot, the NC scale is chosen as $\Lambda = 0.6,~0.8$ and $1.0$~TeV and the orientation angle $\eta = \pi/2$. From the figures corresponding to $\sqrt{s}=0.5,~1.0,~1.5~\rm{TeV}$, we see that at a given machine energy, as we increase $\Lambda$ from $0.6~\rm{TeV}$ to $1.0~\rm{TeV}$, the fluctuation gets diminished and eventually it becomes zero in the $\Lambda \to \infty$ limit (the SM result). The plots show peaks at $t = 0.35~{\rm {T_{day}}}$ and  $0.85~~{\rm {T_{day}}}$ times of the day, where $T_{day}=23.934~{\rm hours}$. For the machine energy $\sqrt{s}=0.5,~1.0,~1.5$, we see that with the increase although the location of several peaks/dips remains unchanged and the fluctuation pattern remains almost same, however it's magnitude at any particular point of time in a day   for a give $\Lambda$ changes largely. 
 However, for $\sqrt{s}=3.0~\rm{TeV}$, we see something different behaviour: there are dips and peaks in the plot corresponding to $\Lambda = 0.6~\rm{TeV}$ and $0.8~\rm{TeV}$ at some $T_{ratio}$, the same $T_{ratio}$ corresponds to dips and peaks for the $\Lambda = 1.0~\rm{TeV}$ plot i.e. they are out-of-phase. Also interestingly, the height of the peak/dip decreases with the decrease in $\Lambda$, contrary to the one found in $\sqrt{s} = 0.5,~1.0$ and $1.5~\rm{TeV}$ cases. 
%Note that the plot corresponding to $\Lambda = 1~\rm{TeV}$ for $\sqrt{s} = 3~\rm{TeV}$ corresponds to the first minimum in Fig. \ref{fig:del_sigma_vs_rts} ((a), the left figure) and whereas the plots corresponding to $\Lambda = 600~\rm{GeV}$ and $800~\rm{GeV}$ (which corresponds to the second minimum in Fig. \ref{fig:del_sigma_vs_rts} for $\Lambda = 600(800)~\rm{GeV}$).
%%%%%%%%%%%%%%%%%%%%%%%%%%%%%%%%%%%%%%%%%%%%%%%%%%%%%%%%%%%%%%%%%%%%%%%%%%%%
%{\pk {However, for $\sqrt{s}=3.0~\rm{TeV}$, we see something different behaviour: there are dips and peaks in the plot corresponding to $\Lambda = 0.6~\rm{TeV}$ and $0.8~\rm{TeV}$ at some $T_{ratio}$, the same $T_{ratio}$ corresponds to dips and peaks for the $\Lambda = 1.0~\rm{TeV}$ plot i.e. they are out-of-phase.   
%Also interestingly, the height of the peak/dip decreases with the decrease in $\Lambda$, contrary to the one found in $\sqrt{s} = 0.5,~1.0$ and $1.5~\rm{TeV}$ cases. $===0000000==>$ We would see similar behaviour in Fig(10-c) if we also include $\Lambda = 0.4, 0.5 TeV$ $=====>$ What I understand, if you keep choosing $\Lambda$ around first minima in Fig(4-a), you'll get same phase. As soon as get into second oscillation, get the phase flipped. Probably related with the comments after eq(8).  }}

%%%%%%%%%%%%%%%%%%%%%%%%
\subsection{Angular distributions in presence of earth rotation}
%%%%%%%%%%%%%%%%%%%%%%%%%
The anisotropy emarged from the breaking of Lorentz invariance as mentioned earlier, persists in the time-averaged (averaged over the side-real day $T_d$) azimuthal distribution of the cross-section $\langle \frac{d\sigma}{d\phi}\rangle_T$ can act as a signature of space-time noncommutativity which is found to be absent in many theories beyond the Standard Model physics.
%%%%%%%%%%%%%%%%%%%%%%
\begin{figure}[htb]
 \centering
\includegraphics[width=2.0in,height=2.0in,keepaspectratio]{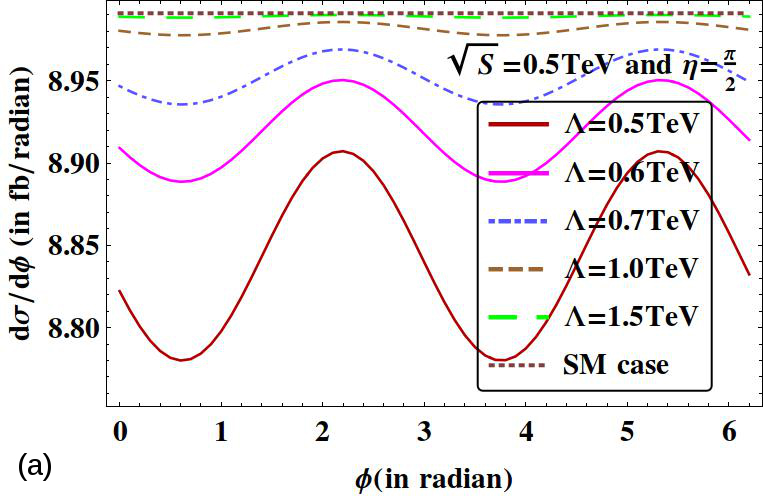}
\includegraphics[width=2.0in,height=2.0in,keepaspectratio]{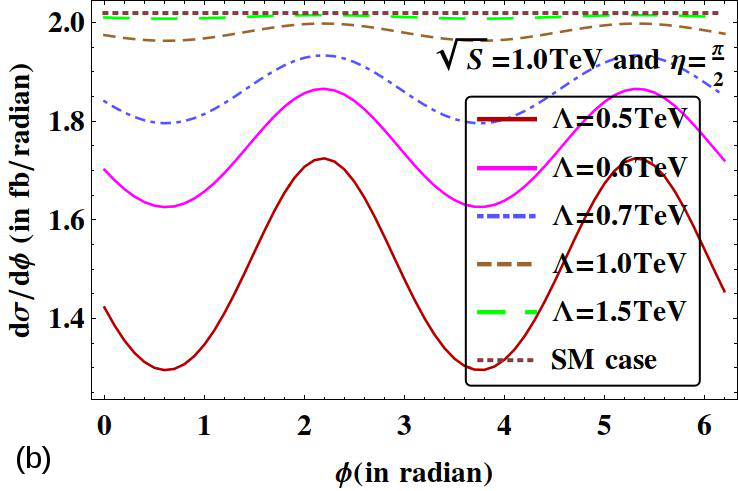}
\includegraphics[width=2.0in,height=2.0in,keepaspectratio]{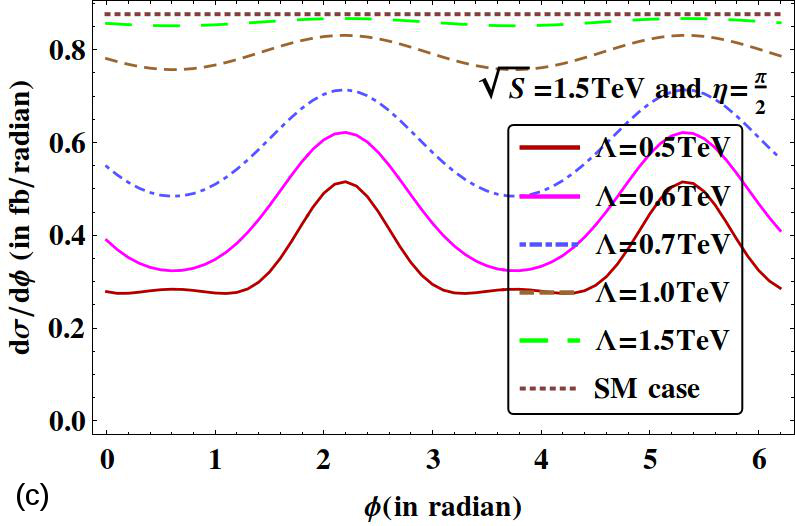}
\includegraphics[width=2.0in,height=2.0in,keepaspectratio]{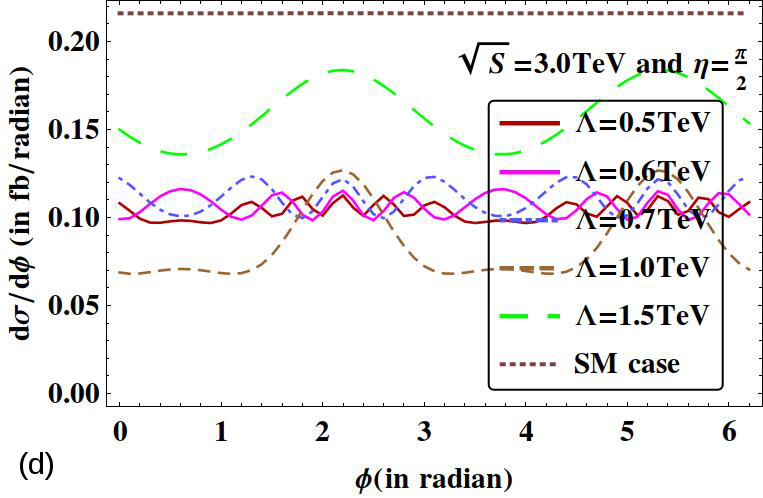}
\caption{\it The time-averaged azimuthal distribution of the cross-section 
$\langle \frac{d\sigma}{d\phi}\rangle_T$ (fb/rad)  is shown as a function of the azimuthal angle $\phi$ (in radian) for $\eta= \pi/2$ and the machine energy $\sqrt{s} = 0.5,~1.0,~1.5$ and $3.0$ TeV (which corresponds to figures a,b,c and d). The different plot in each figure correspond to $\Lambda = 0.5,~0.6,~0.7,~1.0$ and $1.5$ TeV.}
\label{fig:azimuthal_phi}
\end{figure}
%%%%%%%%%%%%%%%%%%%%%%%%%%%%%%%%%%%%%%%%%%%%%%%%%%%%%%

In Fig.~\ref{fig:azimuthal_phi}, we have shown the time averaged azimuthal distribution of the cross-section $\langle \frac{d\sigma}{d\phi}\rangle_T$ corresponding to different values of NC scale $\Lambda = 0.5,~0.6,~0.7,~1.0$ and $1.5~\rm{TeV}$.  The machine energy $\sqrt{s}$ is fixed at $0.5,~1.0,~1.5$ and $3.0~\rm{TeV}$, respectively. Each distribution has maxima(crest) and minima(trough) located at $\phi = 2.2,~5.4~\rm{rad}$ ($0.8,~3.8~\rm{rad}$). At a fixed machine energy (say $\sqrt{s} = 1.0~\rm{TeV}$), the height of the peaks decreases with the increase in $\Lambda$.  For $\eta = \pi/2$, as the machine energy is increased from $0.5~\rm{TeV}$ to $3~\rm{TeV}$, the peaks corresponding to $\Lambda = 0.5,~0.6,~0.7~\rm{TeV}$ located at $\phi = 2.2,~5.4~\rm{rad}$  gets smeared, while the peaks corresponding to $\Lambda = 1.0,~1.5~\rm{TeV}$ still  survives. The fluctuation observed at the crest and trough of the figure corresponding to machine energy $\sqrt{s} = 3~\rm{TeV}$ is due to the fact that $\langle \frac{d\sigma}{d\phi}\rangle_T$ is 
a oscillatory function of $\phi$ and as well as the oscillatory function of $ (\sqrt{s},\Lambda)$. As we see that the behaviour of $\langle \frac{d\sigma}{d\phi}\rangle_T$ at large  energy ($\sqrt{s}$) is dominated by the machine energy at a given $\Lambda$.  
% 

\begin{figure}[htb]
 \centering
\includegraphics[width=1.9in,height=2.9in,keepaspectratio]{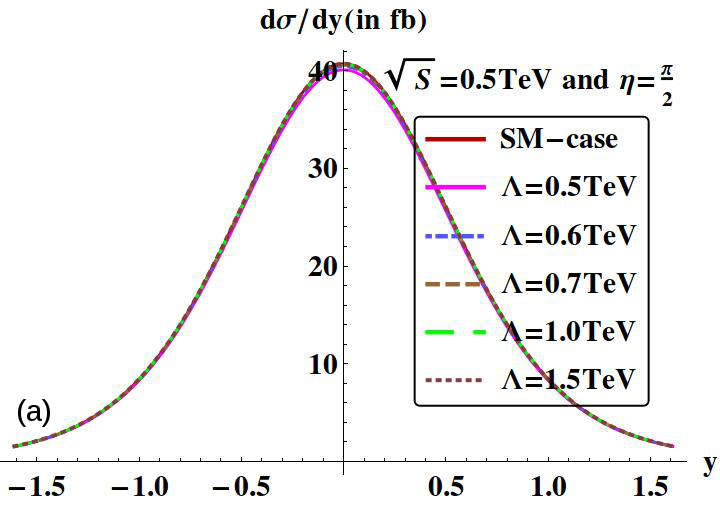}
\includegraphics[width=1.9in,height=2.9in,keepaspectratio]{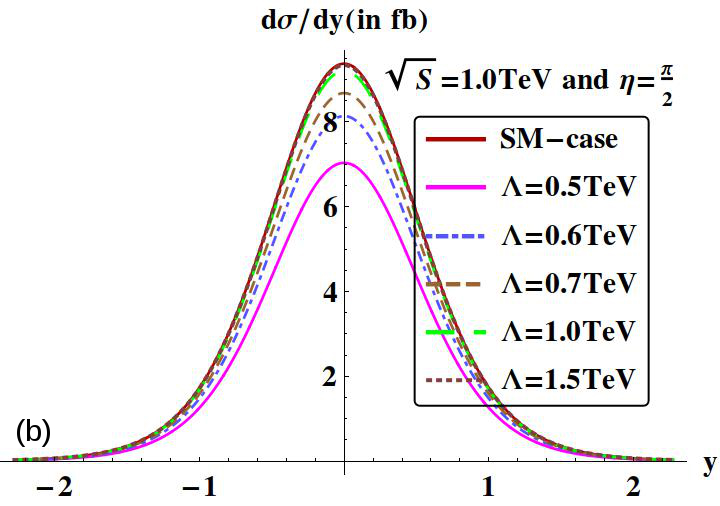}
\includegraphics[width=1.9in,height=2.9in,keepaspectratio]{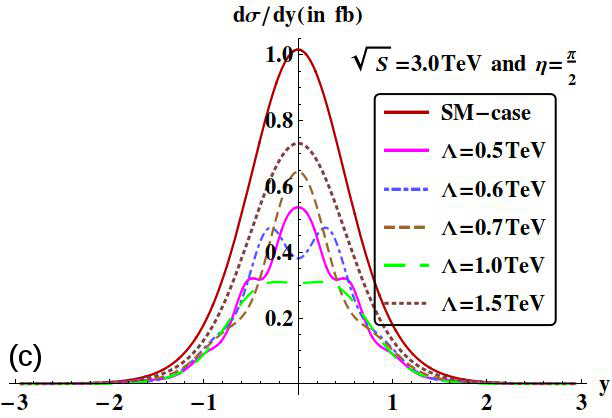}
\caption{\it The rapidity distribution $\langle \frac{d\sigma}{dy} \rangle_T$ (in GeV) is shown plotted as a function of the rapidity $y$ for $\eta = \pi/2$ and the machine energy $\sqrt{s} = 0.5,~1.0$ and $3.0$ TeV (which corresponds to figures a,b and c). The different plot in each figure correspond to $\Lambda = 0.5,~0.6,~0.7,~1.0$ and $1.5$ TeV.}
\label{fig:rapidityLR}
\end{figure}

%%%%%%%%%%%%%%%%%%%%%%%%%%%%%%%%%%%%%%%%%%%%%%%%%%%%%%%%%%%%%%%%%%%%%%%%
In Fig.~\ref{fig:rapidityLR}, we have also shown the time averaged rapidity distribution $\langle \frac{d\sigma}{dy} \rangle_T$ against the rapidity $y$ of the final state particle for different value of the machine energy and $\eta = \pi/2$.
%%%%%%%%%%%%%%%%%%%%%%%%
%%%%%%%%%%%%%%%%%%%%%%
On comparing Fig.~\ref{fig:rapidityLR} and Fig.~\ref{fig:rapidityNR}, we see that the behaviour of the rapidity distributions in both cases(in the presence and absence of the effect due to earth rotation) are almost identical, although the  magnitude differs from each other slightly. 
  
%%%%%%%%%%%%%%%%%%%%%%
%\newpage
%%%%%%%%%%%%%%%%%%%%%%%%%%%%%%%%%%%%%%%%%%%%%%%%%%%%%%%%%%%%%%%%%%%%%
\section{Summary and Conclusion}
\label{sec:conclusion}
In this paper, we  have investigated the  associated Higgs production with $Z$ boson at the future TeV energy linear collider. We did our calculation in the framework of the non-minimal noncommutative standard model(nmNCSM) using the Feynman rules involving all orders of the noncommutative parameters $\Theta_{\mu\nu}$ with (or without) considering the effect of earth rotation. We found that the total cross-section $\sigma(e^- e^+ \to Z H)$ departs significantly from the standard model value as the machine energy starts getting larger than $1.0~\rm{TeV}$ with the NC contribution found to be lower than the SM one. Considering the effect of earth rotation in our analysis, we find from $\langle \Delta \sigma \rangle_T $ vs $\eta$ plot that for $\eta = \pi/2$, $\langle \Delta \sigma \rangle_T $ becomes minimum (yielding $\langle \sigma_{NC} \rangle_T$ to maximum  for $\Lambda \sim 0.6~\rm{TeV}$. The time-averaged NC correction $\langle \Delta \sigma \rangle_T $ and the relative correction $\langle \delta_r \rangle_
T$ is found to be minimum (a trough) (yielding $\langle \sigma_{NC} \rangle_T$ to maximum) at $\sqrt{s} = 1.5~\rm{TeV}$ for the NC scale $\Lambda = 0.5~\rm{TeV}$. The trough depth decreases with the increase in $\Lambda$ and becomes zero as $\Lambda \to \infty$ (the SM value). The diurnal modulation of the NC signal is found to be quite interesting. We plot $\sigma(t) - \langle \sigma \rangle_T$ is plotted as a function of $t/T_{day}$ and is found to have an oscillatory behaviour. At a given energy, the amplitude of oscillation gets damped with the increase in $\Lambda$ and finally it becomes zero in the limit $\Lambda \to \infty$ (the SM limit). The time-averaged azimuthal distribution $\langle \frac{d\sigma}{d\phi} \rangle_T$  against $\phi$  for different $\Lambda$  at different machine energies is found to have a oscillatory behaviour because of the additional terms $p_3\Theta p_4 \sim \left(\frac{\sqrt{s}}{\Lambda}\right)^2~\left(cos \theta + sin \theta \left(sin \phi + cos \phi \right) \right) $ which 
content the tensorial effect from noncommutativity. Note that the distribution is completely flat in the standard model. The distribution shows peak at certain $\phi$ values (similar peaks are observed in the case of no earth rotation)  corresponding to different $\Lambda$ at a machine energy ($\sqrt{s} = 0.5~\rm{TeV}$ to $3.0~\rm{TeV}$), which can be looked at in the linear 
collider experiment and thus test the idea of space-time noncommutativity in near future. 

\begin{center}
{\large {\bf  Acknowledgments}}
\end{center}
The work of P.K.Das is supported in parts by the CSIR Project (Ref. No. 03(1244)12/EMR-II) and the BRNS Project (Ref. No.2011/37P/08/BRNS). 
The authors would like to thank Mr. Atanu Guha for useful discussions.  

%%%%%%%%%%%%%%%%%%%%%%%%%%%%%%%%%%%%%%%%%%%%%%%%%%%%%%%%%%%%%%%%

\end{document}